\documentclass[aps,prd,preprintnumbers,twocolumn,superscriptaddress,nofootinbib]{revtex4-1}
\usepackage{lmodern}
\usepackage{hyperref}
\hypersetup{colorlinks=true,linkcolor=purple,anchorcolor=blue,citecolor=blue, filecolor=blue,urlcolor=purple,bookmarksnumbered=true,
pdfview=FitB
}
\usepackage{mathrsfs}
\usepackage{rsfso}
\usepackage{color}
\usepackage{xcolor}
\usepackage{ulem}
\usepackage{csquotes}
\colorlet{purple1}{blue!70!red}
\colorlet{darkred}{red!50!black}

\usepackage{graphicx}
\usepackage[bf,SL,BF]{subfigure}
\usepackage{psfrag}
\usepackage{color}
\usepackage{amssymb}
\usepackage{amsmath}
\usepackage{epstopdf}
\usepackage{natbib}
\usepackage{amssymb}
\usepackage{amsmath,amssymb}
\usepackage{mathtools}
\usepackage{float}
\usepackage{color}
\usepackage{leftidx}
\usepackage{booktabs}
\usepackage{latexsym}
\usepackage{revsymb}
\usepackage{multirow}
\usepackage{hypcap}
\usepackage{array}
\usepackage[english]{babel}
\usepackage{amsmath}
\usepackage{cleveref}
\def\orcid#1{\kern .08em\href{https://orcid.org/#1}{\includegraphics[keepaspectratio,width=0.7em]{ORCID_iD.png}}}
\newcommand{\be}{\begin{eqnarray}}
	\newcommand{\ee}{\end{eqnarray}}
	
\def\orcid#1{\kern .08em\href{https://orcid.org/#1}{\includegraphics[keepaspectratio,width=0.7em]{ORCID_iD.png}}}

\newcommand{\ba}{\begin{align}}  
\newcommand{\ea}{\end{align}}  

\usepackage{braket}

\usepackage{comment}
\begin{document}

        
        \title{Gravitational form factors of light mesons from Basis Light-Front Quantization}
	
	\author{Amrita~Sain}
    \email{amrita@impcas.ac.cn}
	\affiliation{Institute of Modern Physics, Chinese Academy of Sciences, Lanzhou, Gansu, 730000, China}
	\affiliation{School of Nuclear Physics, University of Chinese Academy of Sciences, Beijing, 100049, China}
	\affiliation{CAS Key Laboratory of High Precision Nuclear Spectroscopy, Institute of Modern Physics, Chinese Academy of Sciences, Lanzhou 730000, China}
 \author{Sreeraj~Nair}
 \email{sreeraj@impcas.ac.cn}
 \affiliation{Institute of Modern Physics, Chinese Academy of Sciences, Lanzhou, Gansu, 730000, China}
 \affiliation{School of Nuclear Physics, University of Chinese Academy of Sciences, Beijing, 100049, China}
 \affiliation{CAS Key Laboratory of High Precision Nuclear Spectroscopy, Institute of Modern Physics, Chinese Academy of Sciences, Lanzhou 730000, China}
 \author{Chandan Mondal}
 \email{mondal@impcas.ac.cn} 
 \affiliation{Institute of Modern Physics, Chinese Academy of Sciences, Lanzhou, Gansu, 730000, China}
 \affiliation{School of Nuclear Physics, University of Chinese Academy of Sciences, Beijing, 100049, China}
 \affiliation{CAS Key Laboratory of High Precision Nuclear Spectroscopy, Institute of Modern Physics, Chinese Academy of Sciences, Lanzhou 730000, China}
 \author{Xingbo Zhao}
 \email{xbzhao@impcas.ac.cn} 
 \affiliation{Institute of Modern Physics, Chinese Academy of Sciences, Lanzhou, Gansu, 730000, China}
 \affiliation{School of Nuclear Physics, University of Chinese Academy of Sciences, Beijing, 100049, China}
 \affiliation{CAS Key Laboratory of High Precision Nuclear Spectroscopy, Institute of Modern Physics, Chinese Academy of Sciences, Lanzhou 730000, China}
 \author{James P. Vary}
 \email{jvary@iastate.edu} 
 \affiliation{Department of Physics and Astronomy, Iowa State University, Ames, Iowa 50011, USA}
 
 \collaboration{BLFQ Collaboration}

	\date{\today}

\begin{abstract}

We compute the gravitational form factors (GFFs) of the pion and kaon using their light-front wave functions within the Basis Light-Front Quantization framework. The wave functions are obtained by solving a light-front effective Hamiltonian that incorporates three-dimensional confinement along with a color-singlet Nambu--Jona-Lasinio interaction between the constituent quark and antiquark. The form factor $A(Q^2)$ is found to be in overall agreement with recent lattice QCD and dispersive results. In contrast, $D(Q^2)$ is enhanced in magnitude at low $Q^2$ relative to both lattice QCD and dispersive determinations. This behavior arises from extracting the $D$-term using transverse components of the QCD energy--momentum tensor, which are more sensitive to the small-$x$ region and to light-front zero-mode effects in the present truncated framework. Using the resulting GFFs, we determine the mass (matter) and mechanical radii of the pion and kaon and analyze their mechanical structure through the corresponding pressure and shear-force distributions.

\end{abstract}

\maketitle

\section{Introduction}\label{intro}
Exploring the internal structure of hadrons, as bound states of quarks and gluons governed by quantum chromodynamics (QCD), remains a central goal of nuclear and particle physics~\cite{Accardi:2012qut,AbdulKhalek:2021gbh,Anderle:2021wcy,Accardi:2023chb,Gross:2022hyw}. A key direction in this effort is to understand how QCD dynamics generates the mechanical properties of hadrons, including their mass, spin, and internal forces~\cite{Polyakov:2018zvc,Burkert:2023wzr}. These properties are encoded in gravitational form factors (GFFs), which are defined through matrix elements of the QCD energy--momentum tensor (EMT) between hadron states.

Although GFFs are of fundamental importance, their direct experimental determination via gravitational interactions is not feasible. Nonetheless, GFFs are related to the second Mellin moments of generalized parton distributions (GPDs)~\cite{Ji:1996nm,Radyushkin:1997ki,Diehl:2003ny} and hard exclusive processes provide an indirect route to constrain EMT matrix elements~\cite{Ji:1996ek,dHose:2016mda,Kumericki:2016ehc,Radyushkin:1996ru,Collins:1996fb}. For the proton, quark and gluon GFFs have been extracted at Jefferson Lab from deeply virtual Compton scattering and near threshold $J/\psi$ photoproduction, respectively~\cite{Burkert:2018bqq,Duran:2022xag,GlueX:2023pev,Guo:2023pqw,Sain:2026ran}.

The pion and kaon, as pseudoscalar Nambu-Goldstone bosons associated with dynamical chiral symmetry breaking, play a special role in low energy QCD. Their EMT structure is therefore of particular interest for understanding how confinement and chiral dynamics shape the internal structure of light mesons. Compared with the proton, experimental information on light meson GFFs remains limited. A phenomenological constraint on pion GFFs has been obtained using generalized distribution amplitudes in two photon production with Belle data~\cite{Kumano:2017lhr,Belle:2015oin}. On the theory side, light meson GFFs have been investigated in a broad range of complementary nonperturbative approaches, including lattice QCD~\cite{Hackett:2023nkr}, continuum bound state methods based on Dyson--Schwinger and Bethe--Salpeter equations~\cite{Xu:2023izo,Yao:2025fnb}, holographic QCD~\cite{Li:2023izn,Fujii:2024rqd,Liu:2025vfe}, the QCD instanton vacuum~\cite{Liu:2024jno,Liu:2024vkj}, meson dominance models~\cite{Broniowski:2024oyk}, and dispersive analyses~\cite{Cao:2025dkv}. This progress provides a timely context for further studies that connect pion and kaon GFFs to spatial mechanical observables within Hamiltonian based frameworks.

In this work, we compute the GFFs of the light pseudoscalar mesons, the pion and kaon, using light-front wave functions (LFWFs) obtained within the Basis Light-Front Quantization (BLFQ) framework~\cite{Vary:2009gt}, restricting the calculation to the valence Fock sector. The approach is supplemented by Nambu--Jona-Lasinio (NJL) interactions. With quarks as the effective degrees of freedom, the NJL model preserves the chiral symmetry of the Lagrangian while allowing for its dynamical breaking~\cite{Klimt:1989pm,Vogl:1989ea,Vogl:1991qt,Klevansky:1992qe}. It therefore provides a natural effective interaction for describing quark dynamics in light mesons. The effective Hamiltonian incorporates transverse confinement motivated by light front holographic QCD~\cite{Brodsky:2014yha}, longitudinal confinement~\cite{Li:2015zda,Li:2017mlw}, and a color singlet NJL interaction~\cite{Klimt:1989pm}. The resulting nonperturbative light front wave functions from BLFQ have been successfully applied to a broad class of pion and kaon observables, including distribution amplitudes, electromagnetic form factors, charge radii, parton distribution functions, GPDs, and pion to photon transition form factors~\cite{Jia:2018ary,Lan:2019vui,Lan:2019rba,Mondal:2021czk,Adhikari:2021jrh}. Building on these established wave functions, we compute the EMT form factors and study the corresponding mechanical properties, including the pressure, shear force, normal force, mass (matter) and mechanical radii for the quark sector in both the pion and kaon.

We organize the paper as follows. In Sec.~\ref{sec:blfq_njl}, we provide a brief overview of the BLFQ NJL formalism for light mesons. In Sec.~\ref{GFFs}, we define the light meson GFFs and outline their extraction from EMT matrix elements. In Sec.~\ref{results}, we present numerical results for the GFFs and analyze the resulting mechanical properties, including pressure, shear, normal force, mass (matter) radius, and mechanical radius. Finally, Sec.~\ref{sumary} summarizes our findings and provides an outlook.

\section{BLFQ--NJL framework for light mesons}
\label{sec:blfq_njl}

In the BLFQ approach, the internal structure of a meson is encoded in its LFWFs, obtained as an eigenstate of an effective light-front Hamiltonian in a truncated Fock space~\cite{Jia:2018ary}. Throughout this work we restrict to the leading valence sector, so that sea-quark and gluon effects are absorbed into effective interactions and parameters. The mass spectrum and LFWFs follow from the light-front eigenvalue problem
\begin{equation}
	H_{\mathrm{eff}}\,\vert \Psi\rangle=M^2\,\vert \Psi\rangle,
	\label{eq:LF_Schrodinger}
\end{equation}
where $M$ is the meson mass and $H_{\mathrm{eff}}$ acts within the $q\bar q$ sector.

Motivated by previous BLFQ applications to light mesons~\cite{Jia:2018ary,Lan:2019vui,Lan:2019rba,Qian:2020utg,Mondal:2021czk,Adhikari:2021jrh}, we employ an effective Hamiltonian consisting of kinetic energy, transverse confinement, longitudinal confinement, and a color-singlet NJL interaction that incorporates chiral dynamics. Explicitly, our effective Hamiltonian is given by
\begin{equation}
	\begin{aligned}
		H_{\mathrm{eff}} =& \frac{\vec{\kappa}^{\perp\,2} + m_q^{2}}{x}
		+ \frac{\vec{\kappa}^{\perp\,2} + m_{\bar q}^{2}}{1-x}
		+ \kappa^{4}\vec{\zeta}^{\perp\,2} \\
		& - \frac{\kappa^{4}}{(m_q + m_{\bar q})^{2}}
		\,\partial_x \big( x(1-x)\,\partial_x \big)
		+ H_{\mathrm{eff}}^{\mathrm{NJL}} .
		\label{eq:Heff}
	\end{aligned}
\end{equation}
Here $x=k^+/P^+$ is the longitudinal momentum fraction carried by the quark, $\vec{\kappa}^\perp$ is the relative transverse momentum conjugate to the transverse separation $\vec r^\perp$, and $m_q$ and $m_{\bar q}$ are effective quark masses. The holographic variable
\begin{equation}
	\vec{\zeta}^{\perp}\equiv \sqrt{x(1-x)}\,\vec{r}^{\perp}
\end{equation}
is used to implement transverse confinement following light-front holographic QCD, with confinement strength $\kappa$~\cite{Brodsky:2014yha}. The longitudinal confining term complements the transverse potential; the derivative
\(
\partial_x f(x,\vec{\zeta}^\perp)
= \partial f(x,\vec{\zeta}^\perp)/{\partial x}|_{\vec{\zeta}^\perp}
\)
is taken at fixed $\vec{\zeta}^\perp$~\cite{Li:2015zda,Li:2017mlw}. The remaining interaction, $H_{\mathrm{eff}}^{\mathrm{NJL}}$, represents local color-singlet four-fermion operators that provide an effective description of chiral dynamics in the valence sector~\cite{Klimt:1989pm,Vogl:1989ea,Vogl:1991qt,Klevansky:1992qe,Jia:2018ary}.

For the charged pion ($u\bar d$), the NJL interaction adopted in the valence sector reads~\cite{Jia:2018ary}
\begin{align}
	H_{\mathrm{NJL},\pi}^{\mathrm{eff}}  &=G_\pi\, \Big\{
	\overline{u}_{\mathrm{u}s_1'}(p_1')u_{\mathrm{u}s_1}(p_1)\,
	\overline{v}_{\mathrm{d}s_2}(p_2)v_{\mathrm{d}s_2'}(p_2') \nonumber\\
	&+ \overline{u}_{\mathrm{u}s_1'}(p_1')\gamma_5 u_{\mathrm{u}s_1}(p_1)\,
	\overline{v}_{\mathrm{d}s_2}(p_2)\gamma_5 v_{\mathrm{d}s_2'}(p_2') \nonumber\\
	&+ 2\,\overline{u}_{\mathrm{u}s_1'}(p_1')\gamma_5 v_{\mathrm{d}s_2'}(p_2')\,
	\overline{v}_{\mathrm{d}s_2}(p_2)\gamma_5 u_{\mathrm{u}s_1}(p_1)
	\Big\},
	\label{eq:H_eff_NJL_pi}
\end{align}
and for the charged kaon ($u\bar s$) it is taken as~\cite{Jia:2018ary}
\begin{align}
	H^{\mathrm{eff}}_{\mathrm{NJL},K}&=G_K\,\Big\{
	- 2\,\overline{u}_{\mathrm{u}s_1'}(p_1') v_{\mathrm{s}s_2'}(p_2')\,
	\overline{v}_{\mathrm{s}s_2}(p_2) u_{\mathrm{u}s_1}(p_1) \nonumber\\
	&+ 2\,\overline{u}_{\mathrm{u}s_1'}(p_1')\gamma_5 v_{\mathrm{s}s_2'}(p_2')\,
	\overline{v}_{\mathrm{s}s_2}(p_2)\gamma_5 u_{\mathrm{u}s_1}(p_1)
	\Big\}.
	\label{eq:H_eff_NJL_K}
\end{align}
These effective interactions follow from the corresponding two- and three-flavor NJL Lagrangians after a Legendre transform and subsequent projection to the valence sector, retaining only the Dirac structures that contribute to the $q\bar q$ LFWFs used here~\cite{Klimt:1989pm,Vogl:1989ea,Vogl:1991qt,Klevansky:1992qe,Jia:2018ary}. Instantaneous contributions associated with the NJL interaction are neglected in Eqs.~\eqref{eq:H_eff_NJL_pi} and~\eqref{eq:H_eff_NJL_K}; explicit expressions for the corresponding matrix elements in the BLFQ basis are given in Ref.~\cite{Jia:2018ary}. In these formulas $u_{\mathrm{f}s}(p)$ and $v_{\mathrm{f}s}(p)$ denote Dirac spinors, with nonitalic subscripts indicating flavor and italic subscripts indicating spin, and $p_1$ ($p_2$) is the quark (antiquark) momentum.

The meson eigenstate in the valence sector is expanded as
\begin{align}
	&\big\vert\Psi(P^+,\vec{P}^\perp)\big\rangle =\sum_{r,s}\int_{0}^{1}\dfrac{dx}{4\pi x(1-x)}\int\dfrac{d\vec{\kappa}^\perp}{(2\pi)^2}\,\nonumber\\
	&\quad\quad\times\,\psi_{rs}(x,\vec{\kappa}^\perp)\, b_r^\dagger(xP^+,\vec{\kappa}^\perp+x\vec{P}^\perp) \nonumber\\
	&\quad\quad\times\,
	d_s^\dagger((1-x)P^+,-\vec{\kappa}^\perp+(1-x)\vec{P}^\perp)\,|0\rangle,\label{eq:Psi_meson_qqbar}
\end{align}
where $\psi_{rs}(x,\vec{\kappa}^{\perp})$ are the momentum-space LFWFs, $d_s^\dagger$ denotes the creation operator for an antiquark with spin $s$, and $b_r^\dagger$ denotes the creation operator for a quark with spin $r$. To solve Eq.~\eqref{eq:LF_Schrodinger} numerically, we expand $\psi_{rs}$ in an orthonormal product basis consisting of a two-dimensional harmonic oscillator (HO) basis in the transverse direction and a Jacobi-polynomial basis in the longitudinal direction~\cite{Vary:2009gt,Li:2015zda,Jia:2018ary},
\begin{align} 
	\psi_{rs}(x,\vec{\kappa}^\perp) &=\sum_{n, m, l}  \langle n, m, l, r, s | \psi\rangle~ \nonumber\\
	&\times\,\phi_{nm}\left(\dfrac{\vec{\kappa}^\perp}{\sqrt{x(1-x)}};b_h\right)\chi_l(x).\label{eq:psi_rs_basis_expansions}
\end{align}
The transverse basis functions are
\begin{align}
	\phi_{nm}\left(\vec{q}^\perp;b_h \right)
	&=\dfrac{1}{b_h}\sqrt{\dfrac{4\pi n!}{(n+|m|)!}}
	\left(\dfrac{|\vec{q}^\perp|}{b_h}\right)^{|m|}
	\exp\!\left(-\dfrac{\vec{q}^{\perp 2}}{2b_h^2}\right)\nonumber\\
	&\quad\times
	L_n^{|m|}\!\left(\dfrac{\vec{q}^{\perp 2}}{b_h^2}\right)\,e^{im\varphi},
	\label{eq:def_phi_nm}
\end{align}
with $\tan\varphi=q^2/q^1$; $b_h$ is the HO basis scale parameter with dimension of mass and $L_n^{|m|}$ is the associated Laguerre polynomial. The longitudinal basis functions are
	\begin{align}
	\chi_l(x;\alpha,\beta)&= \sqrt{4\pi(2l+\alpha+\beta+1)}\nonumber\\
	&\times\,\sqrt{\dfrac{\Gamma(l+1)\Gamma(l+\alpha+\beta+1)}{\Gamma(l+\alpha+1)\Gamma(l+\beta+1)}} \nonumber\\
	&\times\, x^{\beta/2}(1-x)^{\alpha/2}\,P_l^{(\alpha,\beta)}(2x-1),\label{eq:def_chi_l}
\end{align}
where $P_l^{(\alpha,\beta)}$ is the Jacobi polynomial and the parameters
\begin{equation}
	\alpha=\frac{2m_{\bar q}(m_q+m_{\bar q})}{\kappa^2}
	\quad \text{and} \quad
	\beta=\frac{2m_q(m_q+m_{\bar q})}{\kappa^2},
\end{equation}
are fixed by the quark masses and confinement scale~\cite{Li:2015zda,Jia:2018ary}. The BLFQ amplitudes $\langle n,m,l,r,s\vert\psi\rangle$ are obtained by diagonalizing the Hamiltonian matrix in a truncated basis. We impose the truncation
\begin{equation}
	0 \leq n \leq N_{\mathrm{max}},\quad
	-2 \leq m \leq 2,\quad
	0 \leq l \leq L_{\mathrm{max}},
	\label{eq:nmax}
\end{equation}
where $N_{\mathrm{max}}$ controls the transverse resolution and $L_{\mathrm{max}}$ the longitudinal resolution~\cite{Li:2015zda,Jia:2018ary}. The restriction on $m$ reflects the fact that the NJL interaction does not couple to $|m|\ge 3$ states within this construction~\cite{Jia:2018ary}. The LFWFs are normalized as
\begin{align}
	\sum_{r,s}\int_0^1 \frac{dx}{2x(1-x)}
	\int \frac{d^2 \vec{\kappa}^\perp}{(2\pi)^3}\,
	\big|\psi_{rs}(x,\vec{\kappa}^\perp)\big|^2
	=1.
	\label{eq:normalization_LFWF}
\end{align}

\begin{table}[H]
	\caption{Summary of the model parameters~\cite{Jia:2018ary}.}
	\label{tab:model_parameters}
	\centering
	\begin{tabular}{ccc ccc ccc c}
			\hline\hline
		Meson & $N_{\text{max}}$ & $L_{\text{max}}$ & $\kappa$ (MeV) & $m_q$ (MeV) & $m_{\bar q}$ (MeV) & \\
		\colrule
		$\pi$ & 8 & 8--32 & 227 & 337 & 337 &  \\
		$K$ & 8 & 8--32 & 276 & 308 & 445 &  \\
		\botrule
	\end{tabular}
\end{table}

The model parameters, including quark masses, confinement strength $\kappa$, and NJL couplings, are fixed by reproducing the ground-state masses of light pseudoscalar and vector mesons and the charge radii of the $\pi^+$ and $K^+$~\cite{Jia:2018ary}. In the present study we use the parameter sets summarized in Table~\ref{tab:model_parameters}. The resulting LFWFs have been extensively tested in prior applications to light-meson observables, including distribution amplitudes and elastic electromagnetic form factors~\cite{Jia:2018ary}, pion and kaon PDFs and pion induced Drell--Yan processes~\cite{Lan:2019vui,Lan:2019rba}, pion photon transition form factors~\cite{Mondal:2021czk}, and valence-quark GPDs of the pion and kaon~\cite{Adhikari:2021jrh}.

\section{GRAVITATIONAL FORM FACTORS OF PION AND KAON}\label{GFFs}
The internal structure of hadrons can be probed through matrix elements of local operators, most notably the electromagnetic current and the EMT. In the light-front framework, these matrix elements can be expressed exactly in terms of light-front Fock-state wave functions. The GFFs are obtained from matrix elements of the EMT and, equivalently, are related to the second Mellin moments of GPDs.

The gauge-invariant symmetric QCD EMT is given by~\cite{Harindranath:1997kk}
\begin{align}\label{emtqcd}
	T^{\mu \nu}
	&=
	\frac{1}{2}\bar{\psi}\,i\left[\gamma^{\mu}D^{\nu}+\gamma^{\nu}D^{\mu}\right]\psi
	- F^{\mu \lambda a}F_{\lambda a}^{\nu}
	\nonumber\\
	&\quad
	+ \frac{1}{4} g^{\mu \nu} \left(F_{\lambda \sigma a}\right)^2
	- g^{\mu \nu} \bar{\psi}\left(i\gamma^{\lambda}D_{\lambda}-m\right)\psi\,,
\end{align}
where $\psi$ and $A^\mu$ denote the quark and gluon fields, respectively. The non-Abelian field-strength tensor is
\begin{equation}
	F^{\mu \nu}_a
	=
	\partial^{\mu} A^{\nu}_a
	-\partial^{\nu} A^{\mu}_a
	+ g f^{abc} A^{\mu}_b A^\nu_c\,,
\end{equation}
and the covariant derivative is defined as
\begin{equation}
	iD^{\mu}=i\overleftrightarrow{\partial}^{\mu}+gA^{\mu}\,,\nonumber
\end{equation}
with
\begin{equation}
	\alpha\big(i\overleftrightarrow{\partial}^{\mu}\big)\beta
	=
	\frac{i}{2}\alpha\left(\partial^{\mu}\beta\right)
	-\frac{i}{2}\left(\partial^{\mu}\alpha\right)\beta\,.\nonumber
\end{equation}

In the present work, we retain only the fermionic contribution to the EMT. The last term in Eq.~\eqref{emtqcd} vanishes by the equation of motion, so that the fermionic part reduces to
\begin{equation}\label{emtqcdforquark}
	T^{\mu \nu}_q
	=
	\frac{i}{2}\bar{\psi}\,
	\left[\gamma^{\mu}D^{\nu}+\gamma^{\nu}D^{\mu}\right]\psi\,.
\end{equation}

For a spin-zero hadron, the symmetric EMT matrix element is conventionally parametrized in terms of the GFFs $A(Q^2)$ and $D(Q^2)$ as~\cite{Polyakov:2018zvc}
\begin{align}
	&\langle \Psi(P^+,\vec{P}{\,'}^{\perp})|T^{\mu\nu}(0)|\Psi(P^+,\vec{P}^{\perp})\rangle\nonumber\\
	&
	=
	2\bar{P}^{\mu}\bar{P}^{\nu} A(Q^2)
	+\frac{1}{2}\left(q^\mu q^\nu-q^2 g^{\mu\nu}\right)D(Q^2)\,,
	\label{GFF parameterization Eq}
\end{align}
where $\bar{P}^{\mu}=\frac{1}{2}(P'+P)^{\mu}$ is the average four-momentum, $q^\mu=(P'-P)^\mu$ is the momentum transfer, and $Q^2=-q^2$. We work in the symmetric Drell--Yan frame, where $q^+=0$ and $\bar{P}^{\perp}=0$.

The GFF $A_q(Q^2)$ is extracted from the $T_q^{++}$ matrix element, while $D_q(Q^2)$ is obtained from the transverse component $T_q^{12}$. For convenience, we define
\begin{equation}\label{Matrix element of EMT}
	\mathcal{M}^{\mu \nu}
	=
	\langle \Psi(P^+,\vec{P}{\,'}^{\perp})|T_q^{\mu\nu}(0)|\Psi(P^+,\vec{P}^{\perp})\rangle\,.
\end{equation}
The corresponding relations are
\begin{equation}
\begin{aligned}
	\mathcal{M}^{++}
	&=
	2(P^+)^2 A_q(Q^2)\,,
	\\
	\mathcal{M}^{12}
	&=
	\frac{1}{2}q^{(1)}q^{(2)}D_q(Q^2)\,.
	\label{GFF D parameterization Eq}
\end{aligned}
\end{equation}

The quark GFFs can then be written as overlaps of LFWFs,
\begin{equation}
\begin{aligned}
	\label{eq:ForAandD_lfwf_LightMeson}
	A_q(Q^2)
	&=
	\sum_{r,s}
	\int d\Gamma\, x\,
	\psi_{rs}^{*}(x,\vec{\kappa}{\,'}^{\perp})
	\psi_{rs}(x,\vec{\kappa}^{\perp}),
	\\
	q^{(1)}q^{(2)}D_q(Q^2)
	&=
	\sum_{r,s}
	\int d\Gamma\,
	\psi_{rs}^{*}(x,\vec{\kappa}{\,'}^{\perp})
	\frac{\mathcal{O}_{r}}{2x}
	\psi_{rs}(x,\vec{\kappa}^{\perp})\,,
\end{aligned}
\end{equation}
where the integration measure is $d\Gamma \equiv \frac{dx\,d^2\vec{\kappa}^{\perp}}{(2\pi)^3}\,.$
For the struck quark, $\vec{\kappa}{\,'}^{\perp}=\vec{\kappa}^{\perp}+(1-x)\vec{q}^{\,\perp}$, whereas for the spectator, $\vec{\kappa}{\,'}^{\perp}=\vec{\kappa}^{\perp}-x\vec{q}^{\,\perp}$.
The operator $\mathcal{O}_{r}$ entering the expression for $D_q(Q^2)$ is
\begin{equation}\label{eq:operator}
	\mathcal{O}_{r}
	=
	\frac{1}{4}
	\left[
	2\,\mathcal{R}^{(1)}\mathcal{R}^{(2)}
	+ir\left(\mathcal{R}^{(1)}q^{(1)}-\mathcal{R}^{(2)}q^{(2)}\right)
	\right],
\end{equation}
with $\mathcal{R}^{(j)} \equiv 2\kappa^{(j)}+(1-x)q^{(j)}$. Here, $r=1(-1)$ for the struck quark helicity.

Assuming SU(2) flavor symmetry for the pion ($m_u = m_d$), its GFFs satisfy $A_\pi(Q^2) = 2 A_q(Q^2)$ and $D_\pi(Q^2) = 2 D_q(Q^2)$. For the positively charged kaon, flavor symmetry breaking ($m_u \neq m_{\bar{s}}$) results in unequal contributions from the constituent quark and antiquark, giving $A_K(Q^2) = A_q(Q^2) + A_{\bar{s}}(Q^2)$ and $D_K(Q^2) = D_q(Q^2) + D_{\bar{s}}(Q^2)$.




\begin{figure*}[htp]
	\centering
	\includegraphics[width=0.45\textwidth]{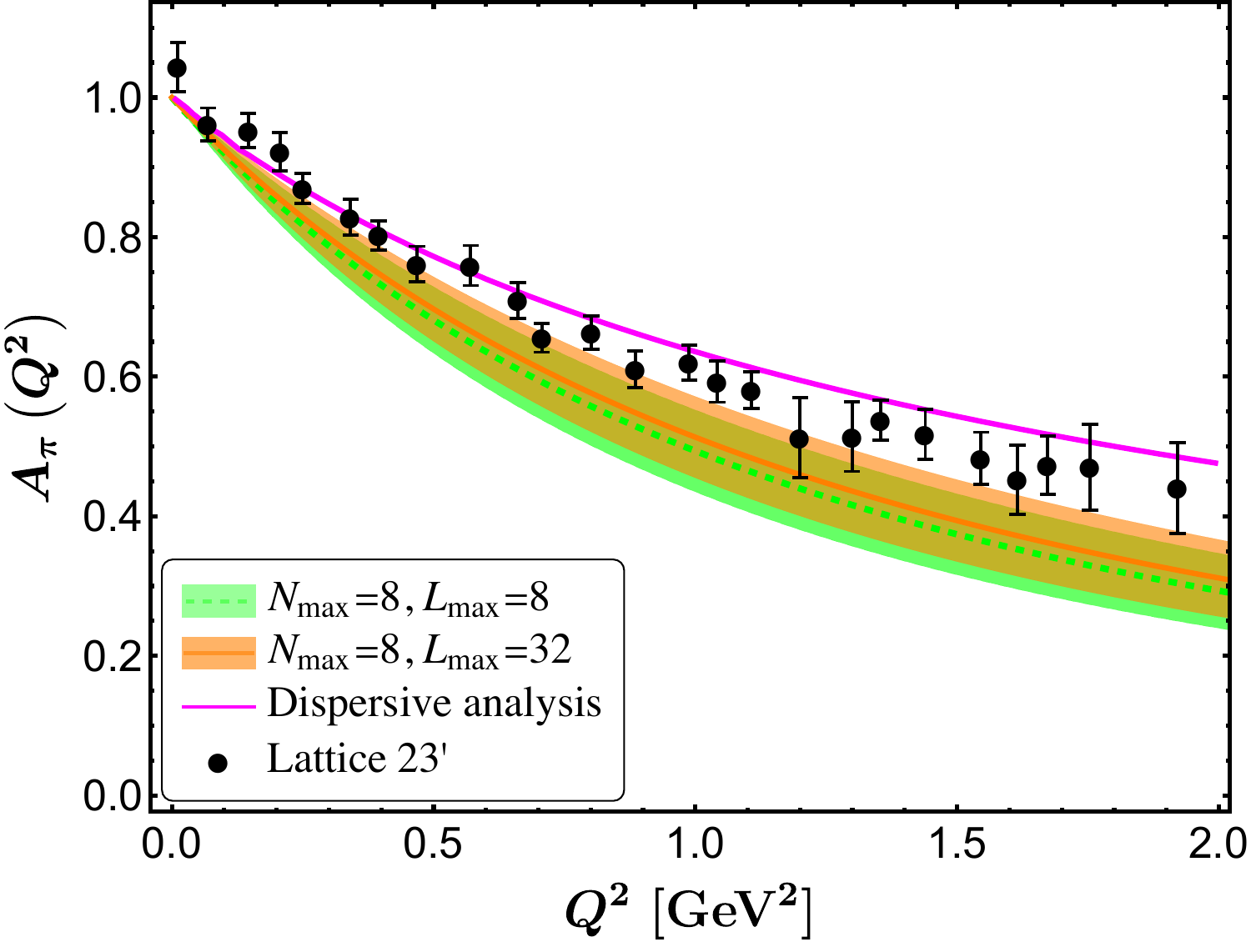}
	\includegraphics[width=0.45\textwidth]{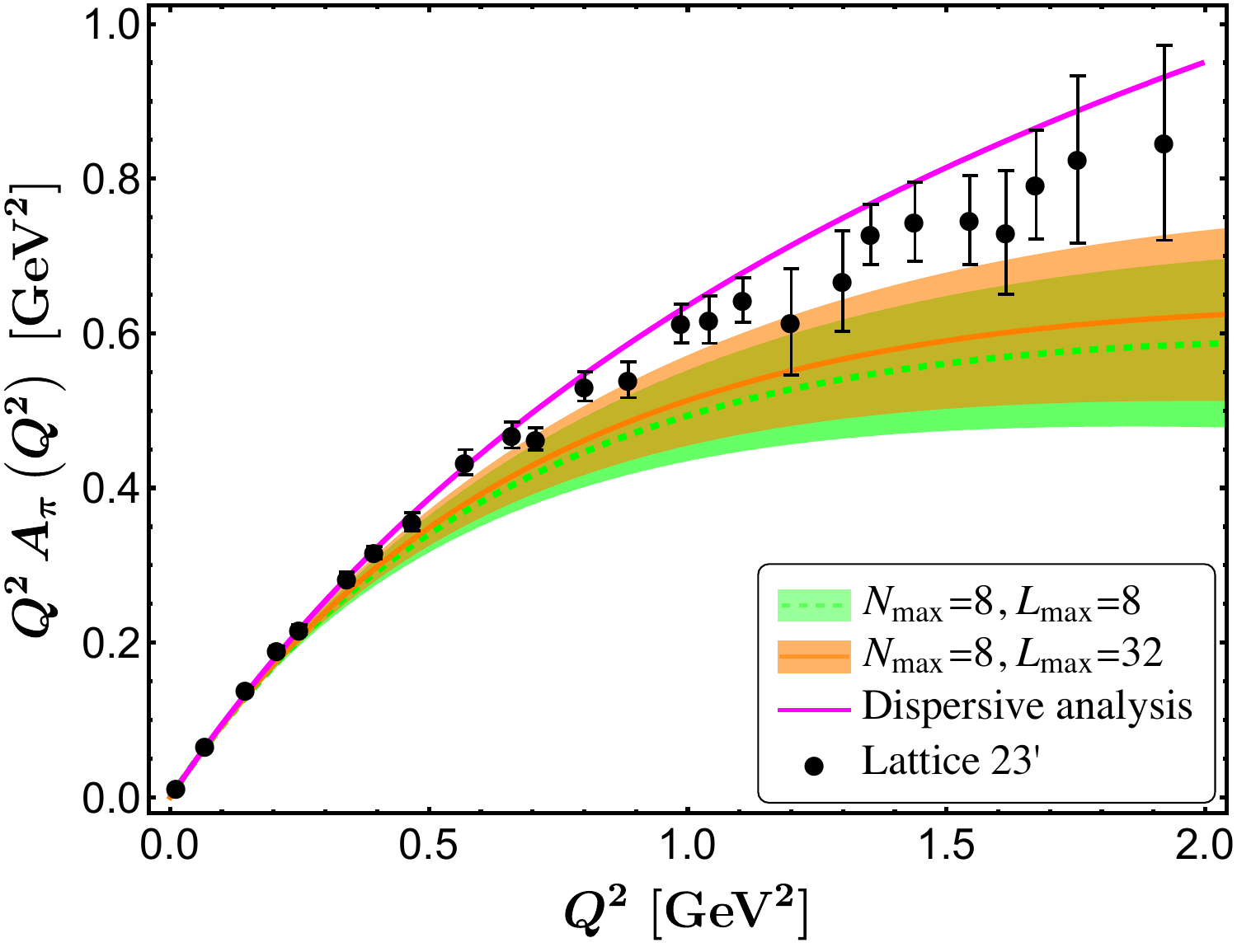}\\
	\includegraphics[width=0.45\textwidth]{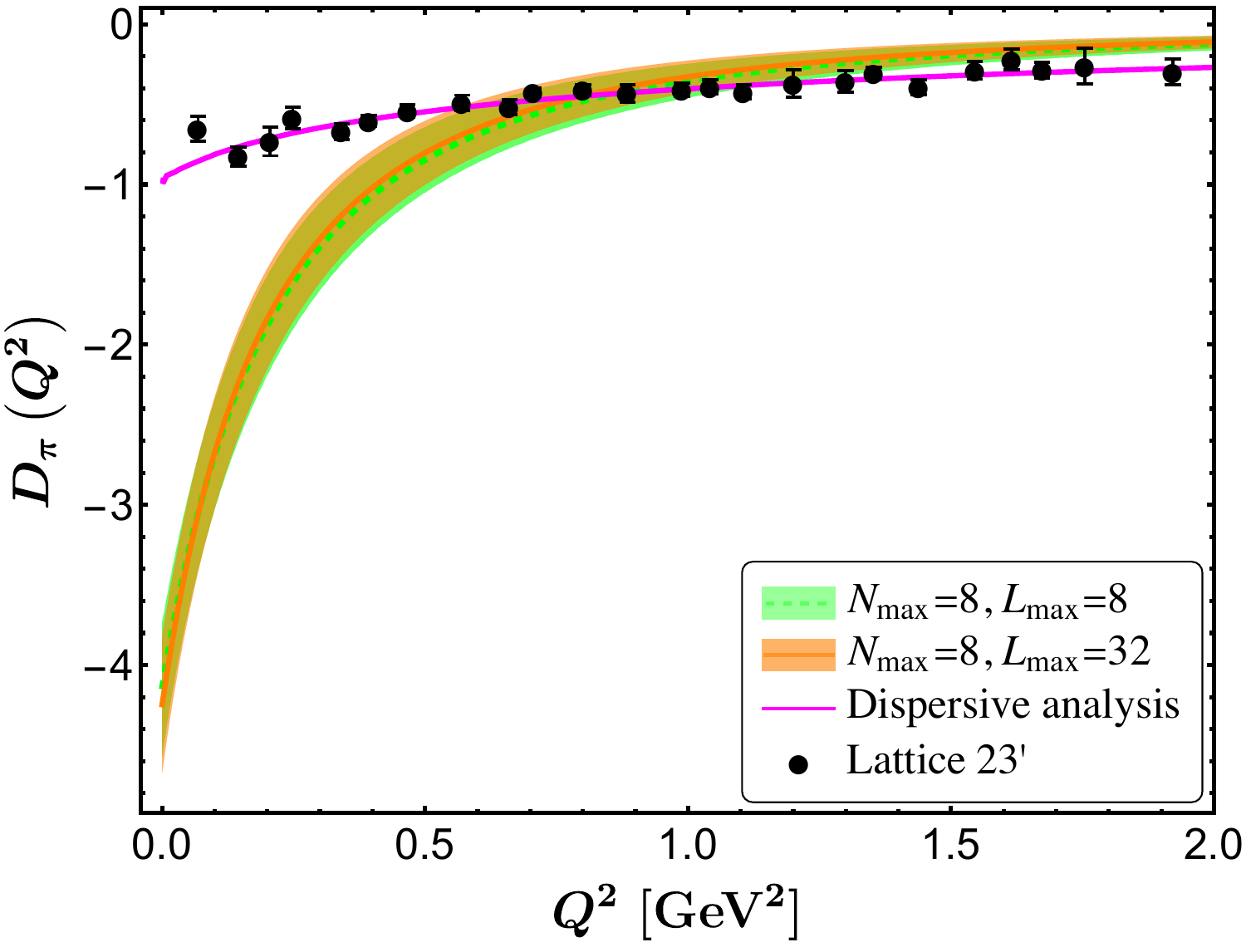}
	\includegraphics[width=0.45\textwidth]{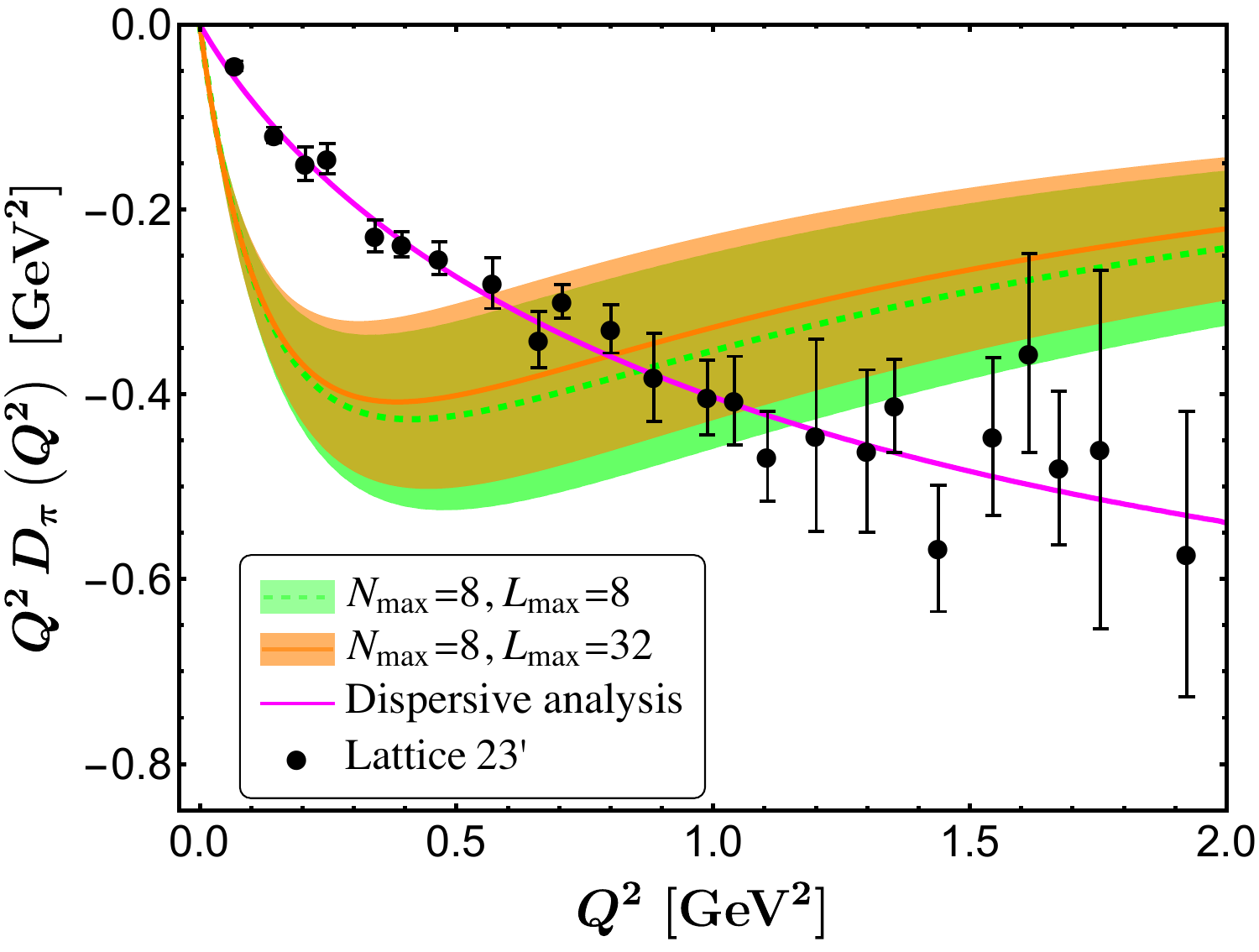}
\caption{Pion GFFs in comparison with lattice QCD results from Ref.~\cite{Hackett:2023nkr} (black points with error bars) and the dispersive analysis of Ref.~\cite{Cao:2025dkv} (magenta curves). The four panels show $A_\pi(Q^2)$ (upper left), $Q^2 A_\pi(Q^2)$ (upper right), $D_\pi(Q^2)$ (lower left), and $Q^2 D_\pi(Q^2)$ (lower right). The green and orange bands represent the BLFQ results for $N_{\max}=8$ with $L_{\max}=8$ and $L_{\max}=32$, respectively. The band widths indicate the uncertainty induced by a $10\%$ variation of the fitted model parameters.}
	\label{fig:pion_GFFs} 
\end{figure*}

\begin{figure*}[htp]
	\centering
	\includegraphics[width=0.45\textwidth]{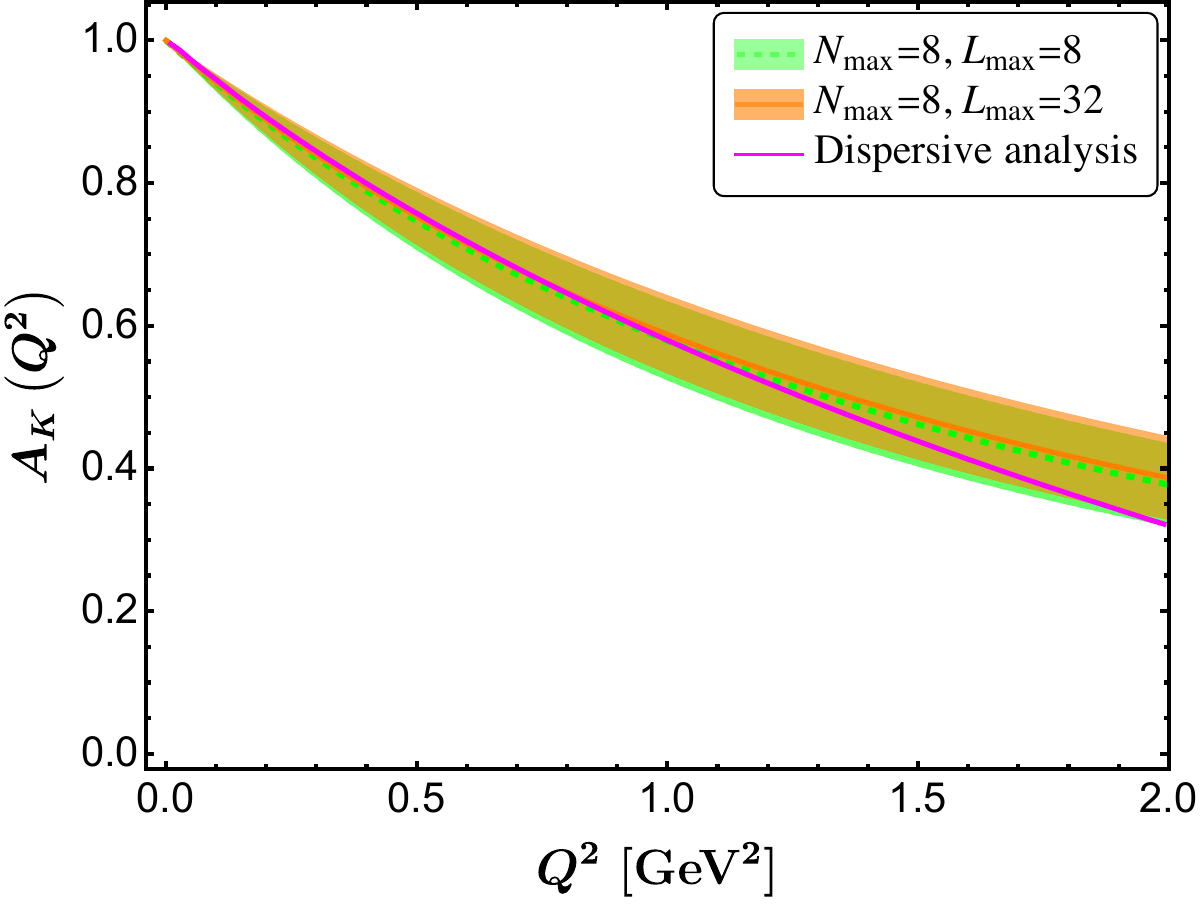}
	\includegraphics[width=0.45\textwidth]{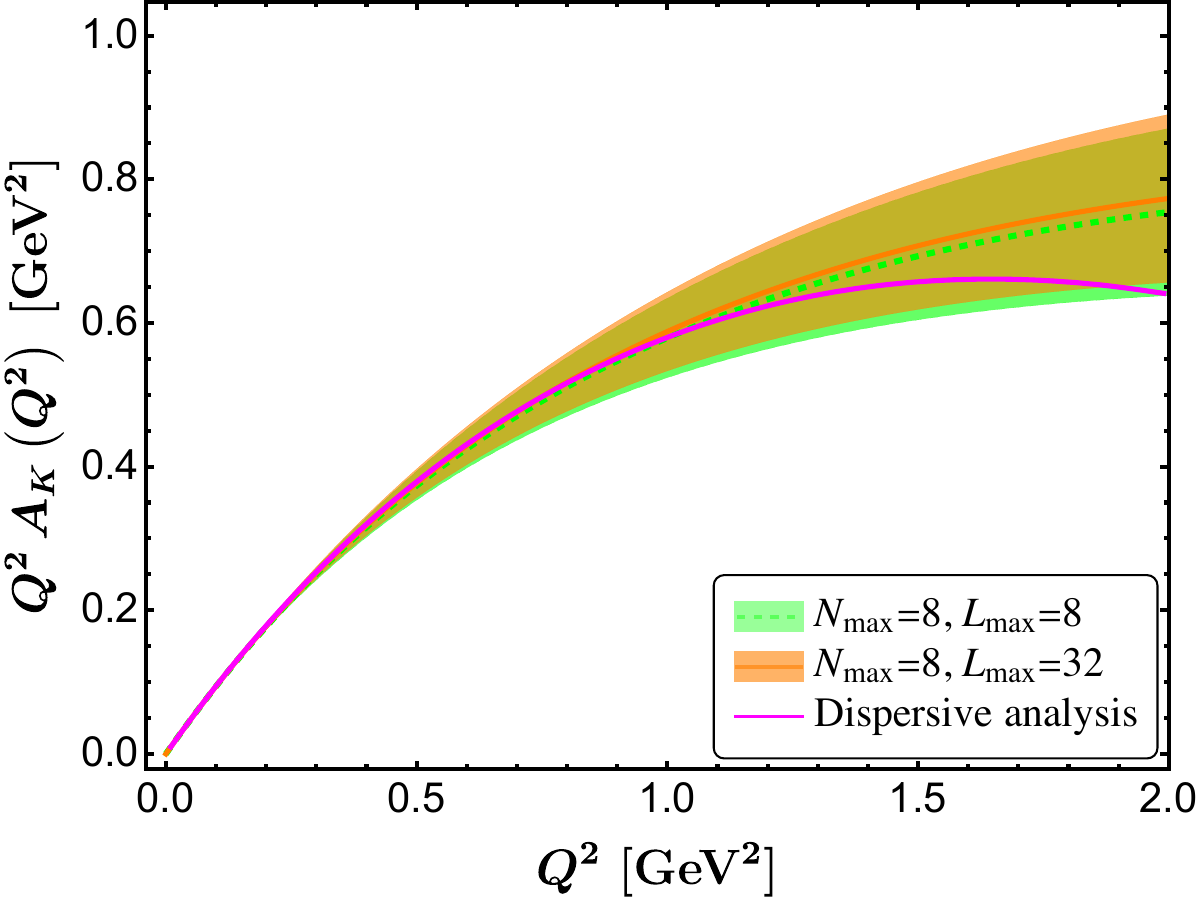}\\
	\includegraphics[width=0.45\textwidth]{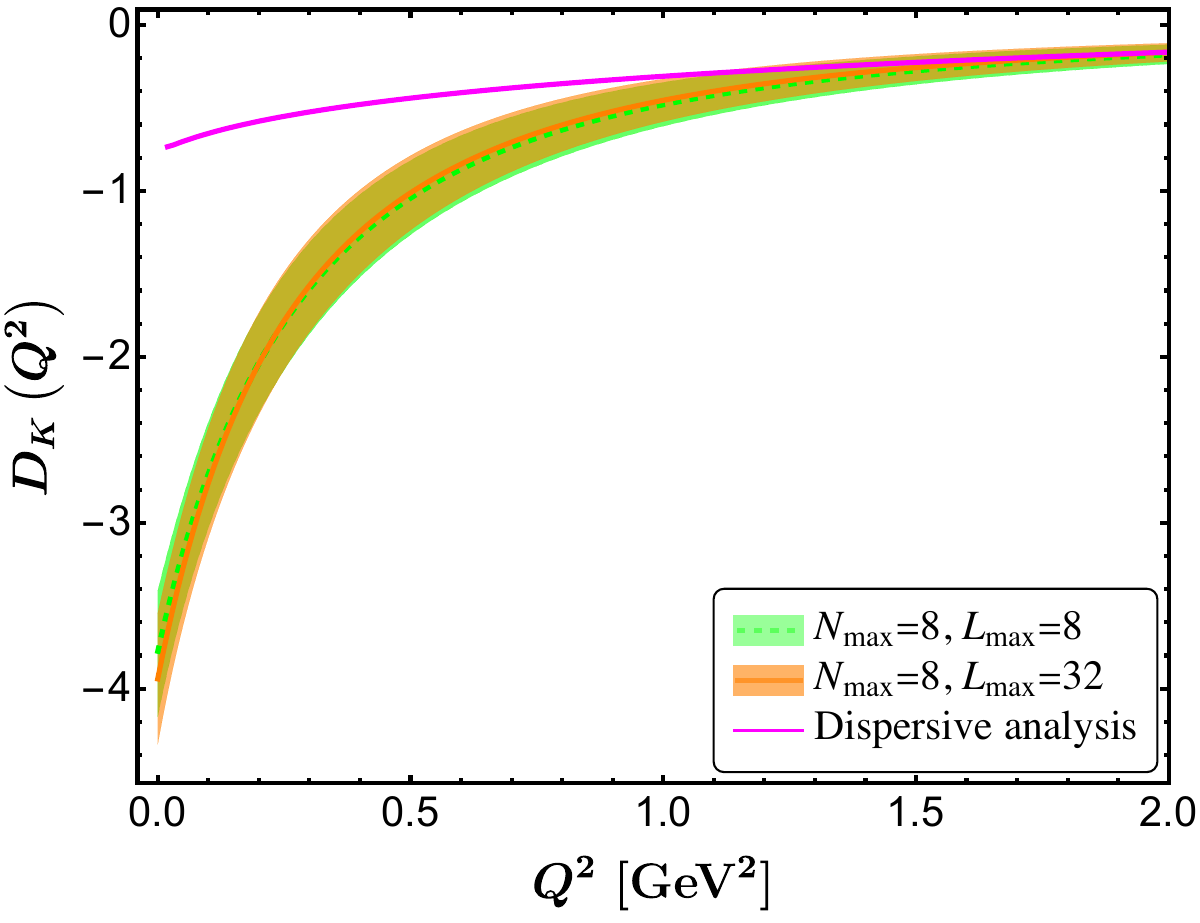}
	\includegraphics[width=0.45\textwidth]{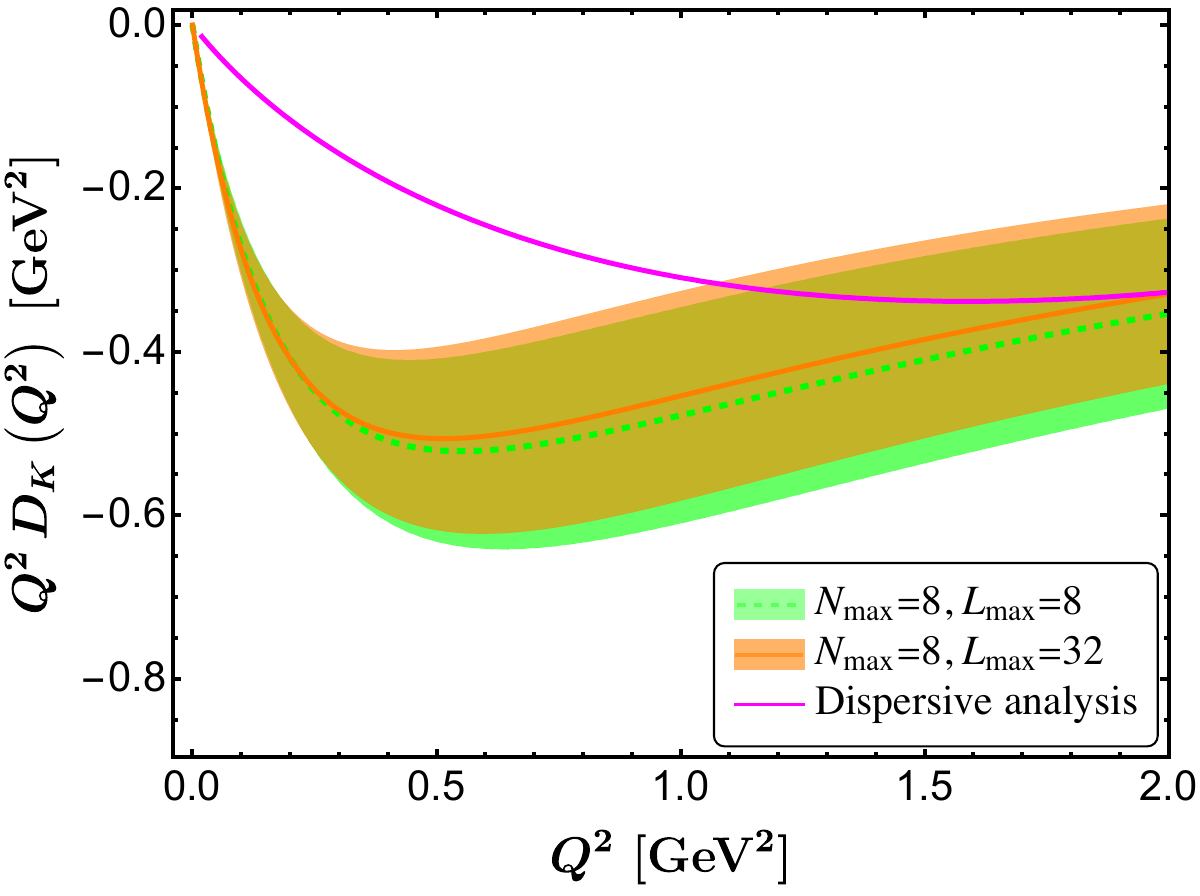}
\caption{Kaon gravitational form factors in comparison with the dispersive analysis of Ref.~\cite{Cao:2025dkv} (magenta curves). The four panels show $A_K(Q^2)$ (upper left), $Q^2 A_K(Q^2)$ (upper right), $D_K(Q^2)$ (lower left), and $Q^2 D_K(Q^2)$ (lower right). The green and orange bands represent the BLFQ results for $N_{\max}=8$ with $L_{\max}=8$ and $L_{\max}=32$, respectively. The band widths indicate the uncertainty induced by a $10\%$ variation of the fitted model parameters.}
	\label{fig:kaon_GFFs} 
\end{figure*}

\begin{figure*}[htp]
	\centering
	\includegraphics[width=0.45\textwidth]{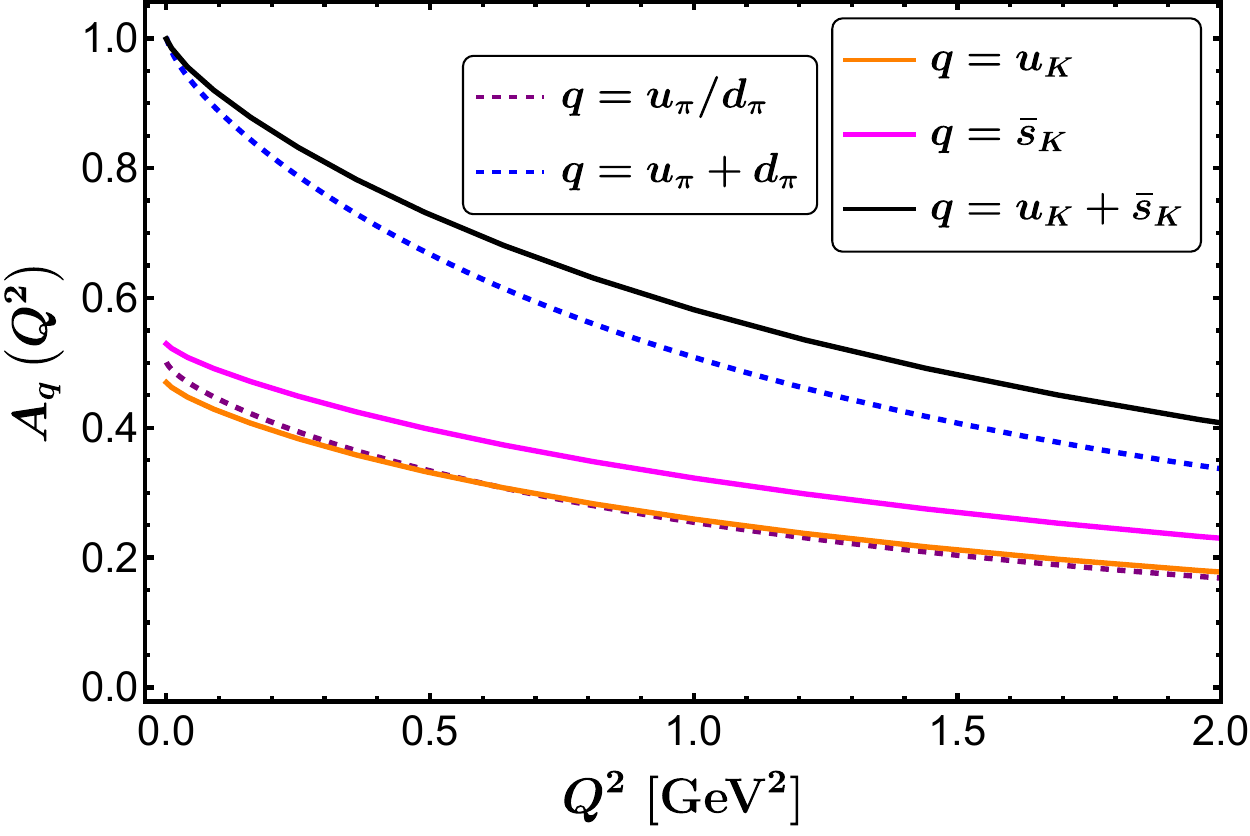}
	\includegraphics[width=0.45\textwidth]{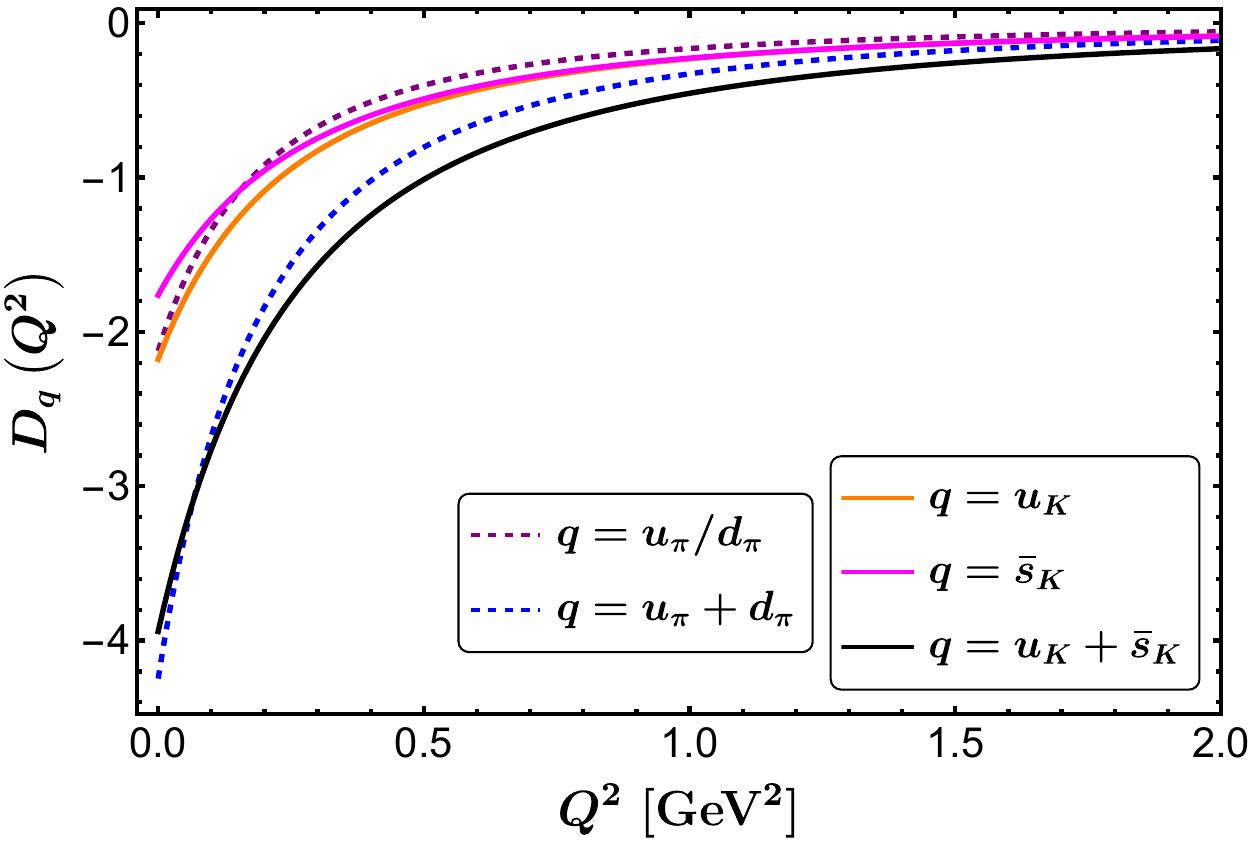}
\caption{Pion and kaon gravitational form factors together with their quark flavor decompositions. The left panel displays $A(Q^2)$, while the right panel displays $D(Q^2)$. The solid black curves correspond to the kaon form factors, $A_K(Q^2)$ and $D_K(Q^2)$, and the blue dashed curves show the pion form factors, $A_\pi(Q^2)$ and $D_\pi(Q^2)$. The orange and magenta curves represent the $u$- and $\bar{s}$-quark contributions in the kaon, respectively. In the pion, the quark and antiquark contributions are equal and are represented by the purple dashed curves.}
	\label{fig:GFFs_partons} 
\end{figure*}

\section{NUMERICAL RESULTS}\label{results}
The LFWFs of the pion and kaon, obtained within the BLFQ--NJL framework~\cite{Jia:2018ary} and briefly reviewed in Sec.~\ref{sec:blfq_njl}, are used in Eq.~\eqref{eq:ForAandD_lfwf_LightMeson} to calculate the meson GFFs. In the numerical analysis, we employ the valence wave functions from Eq.~\eqref{eq:psi_rs_basis_expansions} with basis truncations $N_\text{max} = 8$ and $L_\text{max} = \{8, 32\}$ with the HO scale parameter $b_h$ set to be identical to the confining strength $\kappa$. The model parameters are listed in Table~\ref{tab:model_parameters}.

A technical issue arises for the extraction of the $D$-term at small momentum transfer. In the overlap representation of Eq.~\eqref{eq:ForAandD_lfwf_LightMeson}, the integrand for $D_q(Q^2)$ contains an explicit factor of $1/x$, which enhances the sensitivity to the small-$x$ region. In a light-front calculation restricted to the leading valence sector, this is precisely the kinematic region where omitted zero-mode and higher-Fock-sector contributions can become important. As a result, the low-$Q^2$ behavior of $D(Q^2)$ is much less stable than that of $A(Q^2)$ and develops an enhanced magnitude near $Q^2=0$. A similar difficulty has recently been emphasized in Ref.~\cite{Choi:2025rto}, where the treatment of the pion $D$-term also required special care because of light-front zero-mode sensitivity.

To characterize the low-$Q^2$ behavior of the form factors, we employ the dipole parametrization
\begin{equation}
\begin{aligned}
	\label{dipole_eq}
	A(Q^2) &= \frac{1}{\left(1+\frac{Q^2}{m_A^2}\right)^2},  \\
	D(Q^2) &= \frac{D(0)}{\left(1+\frac{Q^2}{m_D^2}\right)^2}.
\end{aligned}
\end{equation}
The corresponding fit parameters are listed in Table~\ref{tab:dipole_fit}. This parametrization provides a convenient representation of the form-factor behavior in the low-momentum-transfer region. A $10\%$ uncertainty is assigned to the fitting parameters.

\begin{table}[H]
	\caption{Dipole fit parameters for the GFFs.}
	\label{tab:dipole_fit}
	\centering
	\begin{tabular}{lccccc}
		\hline\hline
		Meson & $N_{\text{max}}$ & $L_{\text{max}}$ & $D(0)$ & $m_D$ (GeV) & $m_A$ (GeV) \\
		\hline
		$\pi$ & 8 & 8  & $-4.151$ & $0.642$ & $1.537$ \\
		$\pi$ & 8 & 32 & $-4.247$ & $0.620$ & $1.589$ \\
		$K$   & 8 & 8  & $-3.790$ & $0.742$ & $1.784$ \\
		$K$   & 8 & 32 & $-3.941$ & $0.717$ & $1.814$ \\
		\hline\hline
	\end{tabular}
\end{table}

The pion GFFs $A_\pi(Q^2)$ and $D_\pi(Q^2)$ are shown in Fig.~\ref{fig:pion_GFFs} for the two basis truncations $\{N_\text{max},L_\text{max}\}=\{8,8\}$ and $\{8,32\}$, together with lattice QCD results~\cite{Hackett:2023nkr} and the dispersive analysis~\cite{Cao:2025dkv}. The form factor $A_\pi(Q^2)$ decreases monotonically with increasing $Q^2$. Over the full displayed range, our results remain below the dispersive curve but follow the overall trend of both the lattice QCD and dispersive determinations. The difference between the two basis truncations is relatively small, indicating that $A_\pi(Q^2)$ is fairly stable under the present change of basis size. This behavior is also reflected in the scaled quantity $Q^2A_\pi(Q^2)$, which rises from the origin and then begins to flatten at larger $Q^2$, consistent with the onset of an approximate $1/Q^2$ falloff in the explored momentum-transfer range.

The behavior of $D_\pi(Q^2)$ is qualitatively different from that of $A_\pi(Q^2)$. In the present calculation, $D_\pi(Q^2)$ remains negative over the full displayed range, consistent with the usual interpretation of a negative $D$-term as a signal of mechanical stability. At low $Q^2$, however, its magnitude is substantially enhanced, leading to an extrapolated value $D_\pi(0)\sim -4.25$ for $\{N_\text{max},L_\text{max}\}=\{8,32\}$. This is much larger in magnitude than the values determined by lattice QCD, $D_\pi(0)\approx -1$~\cite{Hackett:2023nkr}, and by chiral perturbation theory, $D_\pi(0)\approx -0.96$~\cite{Donoghue:1991qv}. As $Q^2$ increases, our results move closer to both the lattice QCD and dispersive curves. The two basis truncations nevertheless lead to the same qualitative behavior for $D_\pi(Q^2)$ over the full displayed range.

The kaon GFFs $A_K(Q^2)$ and $D_K(Q^2)$ are shown in Fig.~\ref{fig:kaon_GFFs}. The form factor $A_K(Q^2)$ behaves similarly to the pion case. Our results decrease smoothly with increasing $Q^2$, remain close to the dispersive curve over the displayed range, and exhibit minimal sensitivity to the basis truncation. The scaled quantity $Q^2 A_K(Q^2)$ also shows a gradual flattening at larger $Q^2$, consistent with an approximate $1/Q^2$ falloff in the explored region.

The kaon $D$-term remains negative throughout the displayed range, while its magnitude is enhanced at low $Q^2$, leading to an extrapolated value $D_K(0)\sim -3.94$ for $\{N_\text{max},L_\text{max}\}=\{8,32\}$. Although direct lattice QCD results for the kaon $D$-term are not yet available, chiral effective theory and dispersive analyses suggest a considerably smaller magnitude. The value inferred from Ref.~\cite{Cao:2025dkv} at $Q^2=0$ is, in particular, much smaller in magnitude than the present result. As $Q^2$ increases, our results move closer to the dispersive curve. 

The low-$Q^2$ enhancement of $D_\pi(Q^2)$ and $D_K(Q^2)$ should therefore be regarded as a model-dependent feature of the present BLFQ--NJL calculation rather than as a firm quantitative prediction. In the overlap representation, the extraction of the $D$-term is more sensitive to the small-$x$ region, where effects beyond the present valence-sector treatment may become important. A quantitative reconciliation with lattice QCD, dispersive analyses, and chiral effective theory will likely require an explicit treatment of such contributions and, more generally, an extension of the model space beyond the leading Fock sector. In the present work, the dipole parametrization is used only as a practical representation of the low-$Q^2$ behavior.

Figure~\ref{fig:GFFs_partons} shows the individual quark-flavor contributions to the meson GFFs. For the pion, the $u$- and $d$-quark contributions to both $A_\pi(Q^2)$ and $D_\pi(Q^2)$ are identical because $m_u=m_d$. For the kaon, by contrast, $m_u\neq m_{\bar s}$, and the $u$- and $\bar s$-quark contributions therefore differ for both form factors.

A comment is in order regarding the form factor $\bar{c}(Q^2)$. When the EMT is parametrized at the parton level, an additional GFF $\bar{c}(Q^2)$ appears~\cite{Xing:2022mvk,Sultan:2024hep}. In the present calculation, we obtain $\bar{c}(0)=0$ using $T^{+-}$, as mentioned in Refs.~\cite{Hu:2024edc,Xu:2024hfx}, for both the pion and the kaon, whereas $\bar{c}(Q^2)$ becomes nonzero at finite momentum transfer. A similar nonzero $\bar{c}(Q^2)$ at finite momentum transfer was also found in the charmonium analysis of Ref.~\cite{Hu:2024edc}.


\begin{figure*}[htp]
	\centering
	\includegraphics[width=0.45\textwidth]{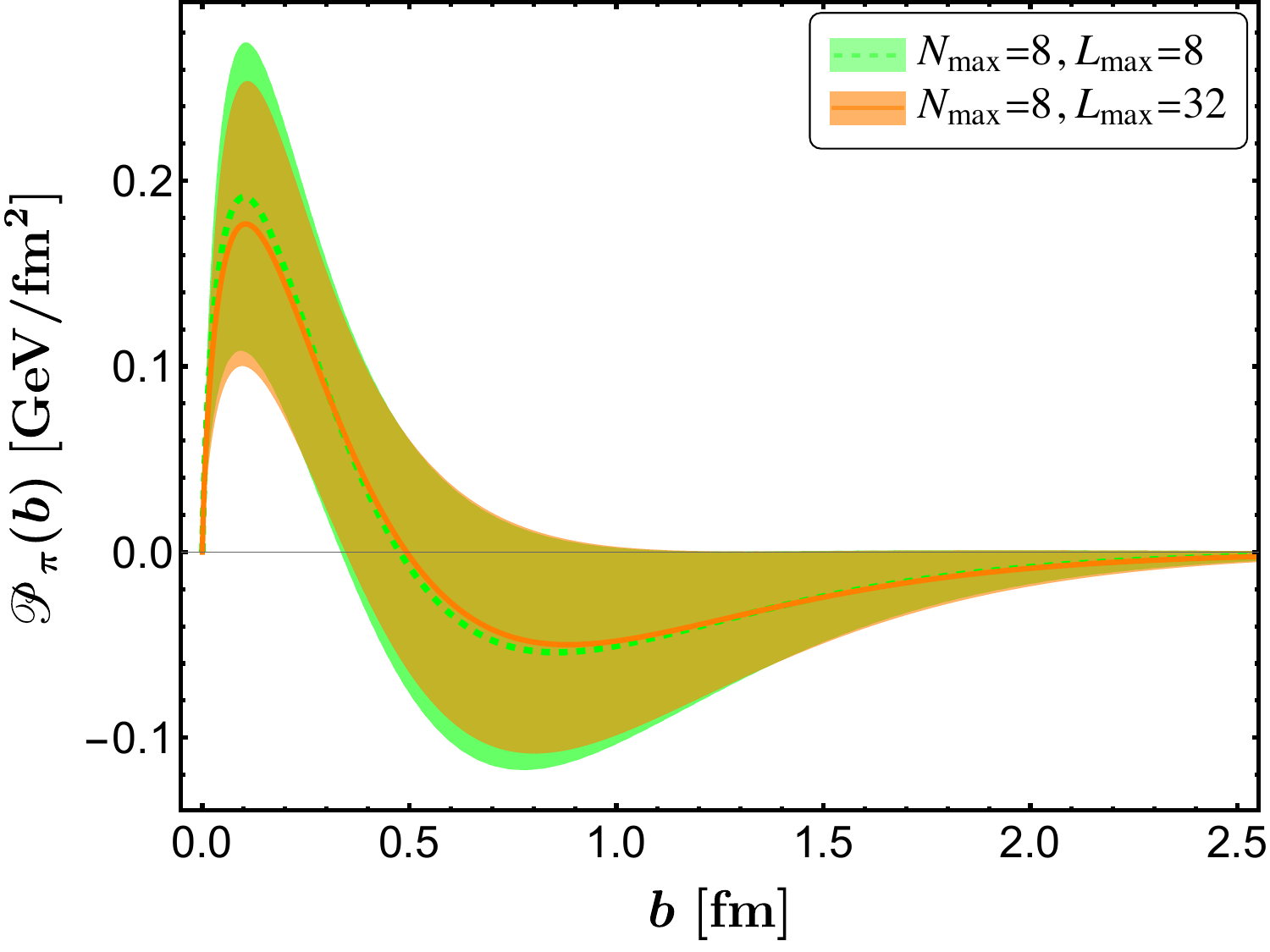}
	\includegraphics[width=0.45\textwidth]{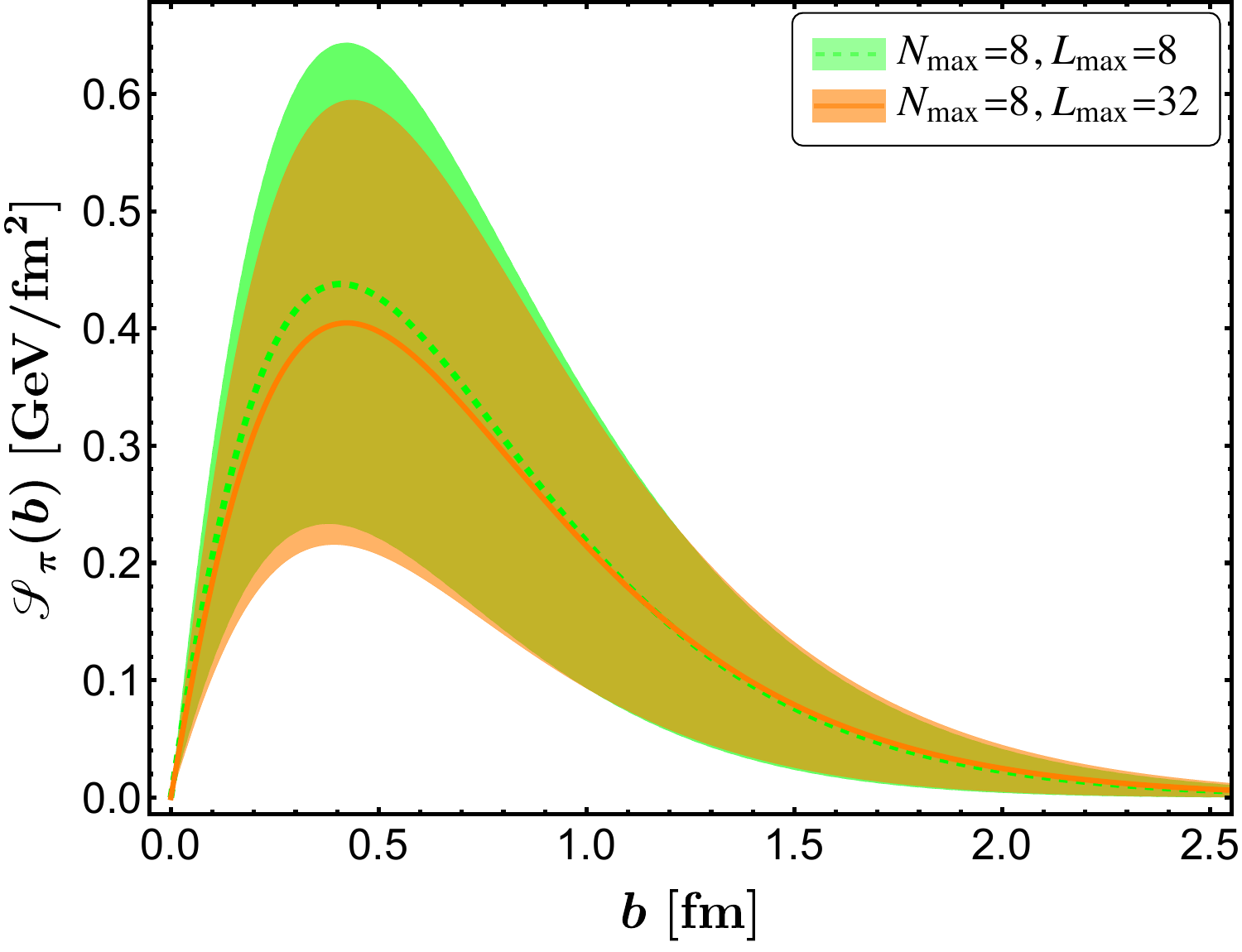}
	\includegraphics[width=0.45\textwidth]{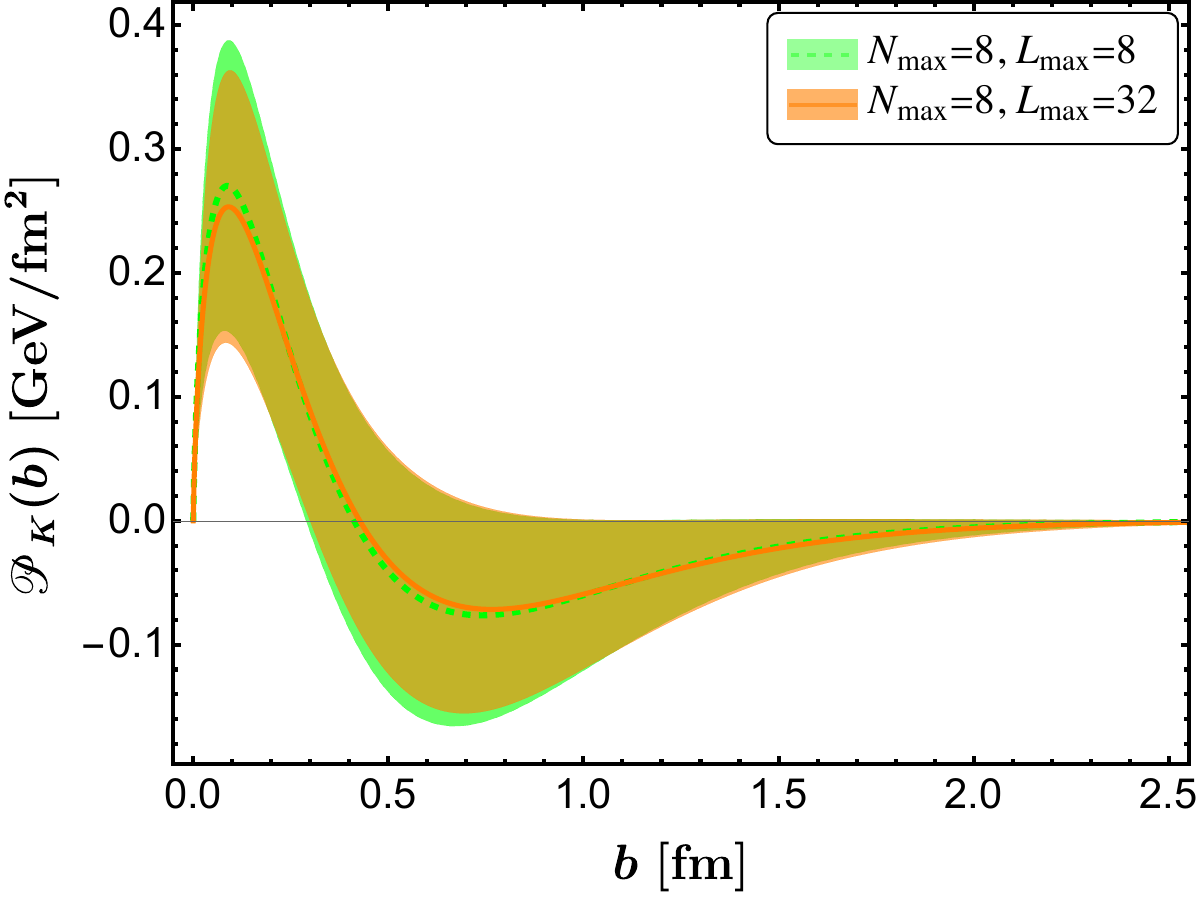}
	\includegraphics[width=0.45\textwidth]{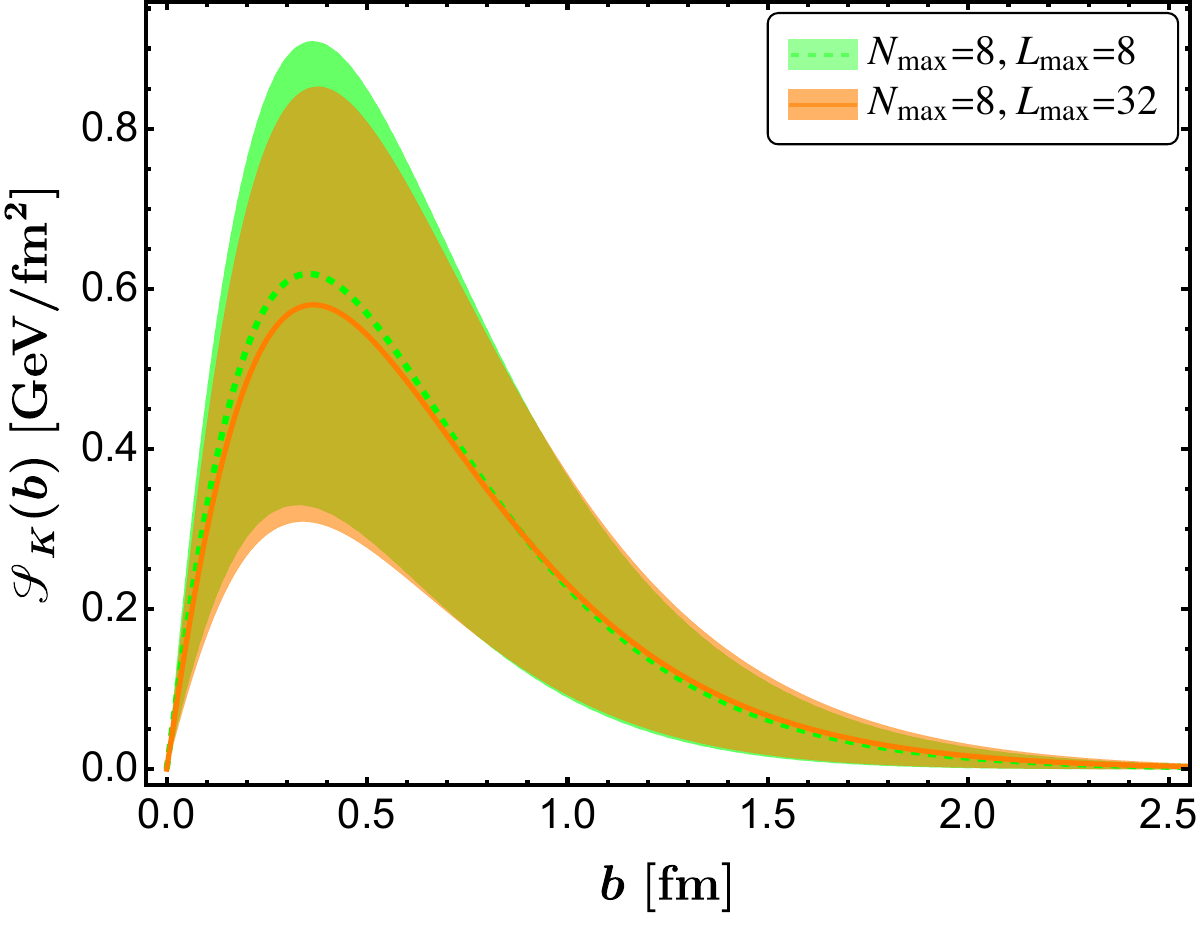}
\caption{Weighted transverse pressure and shear distributions of the pion and kaon. The four panels display $\mathcal{P}_\pi(b)$ (upper left), $\mathcal{S}_\pi(b)$ (upper right), $\mathcal{P}_K(b)$ (lower left), and $\mathcal{S}_K(b)$ (lower right). The green and orange bands represent the BLFQ results for $N_{\max}=8$ with $L_{\max}=8$ and $L_{\max}=32$, respectively. The band widths indicate the uncertainty associated with a $10\%$ variation of the fitted model parameters.}
	\label{fig:mech_densities} 
\end{figure*}

\begin{figure*}[htp]
	\centering
	\includegraphics[width=0.45\textwidth]{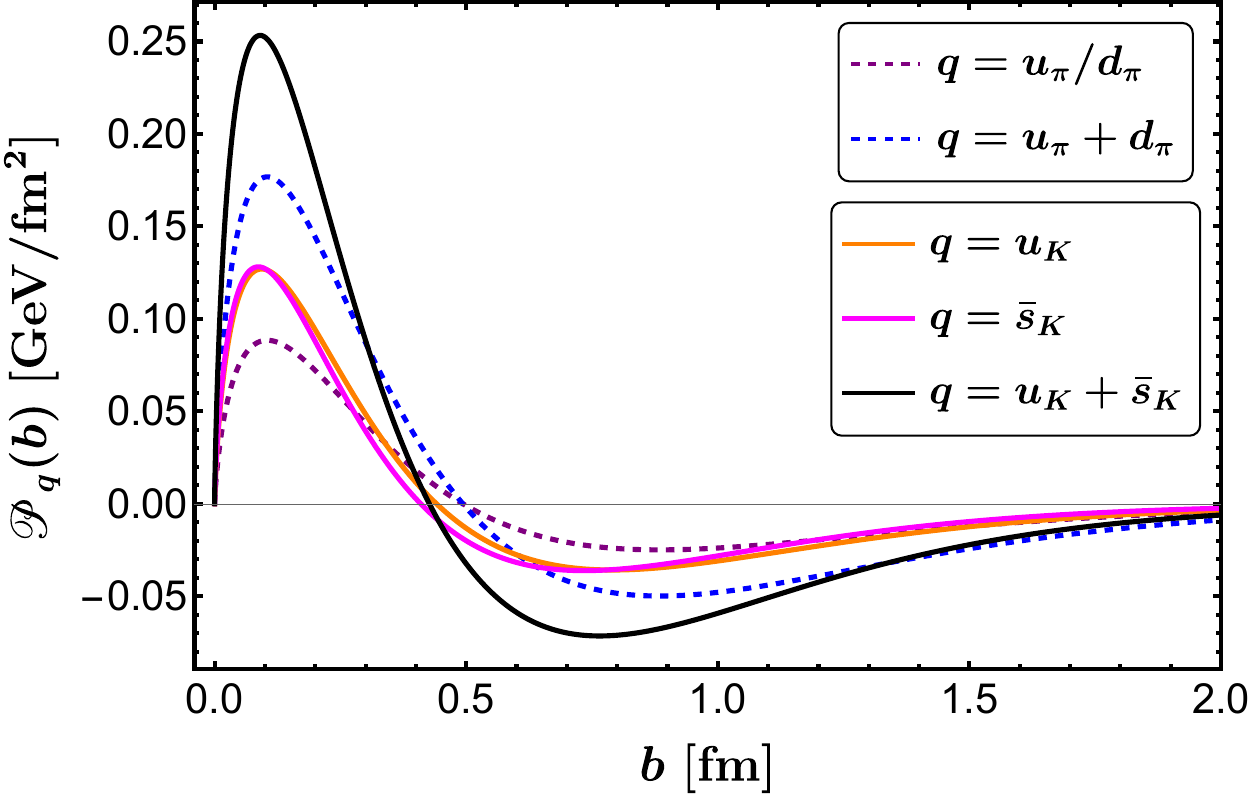}
	\includegraphics[width=0.45\textwidth]{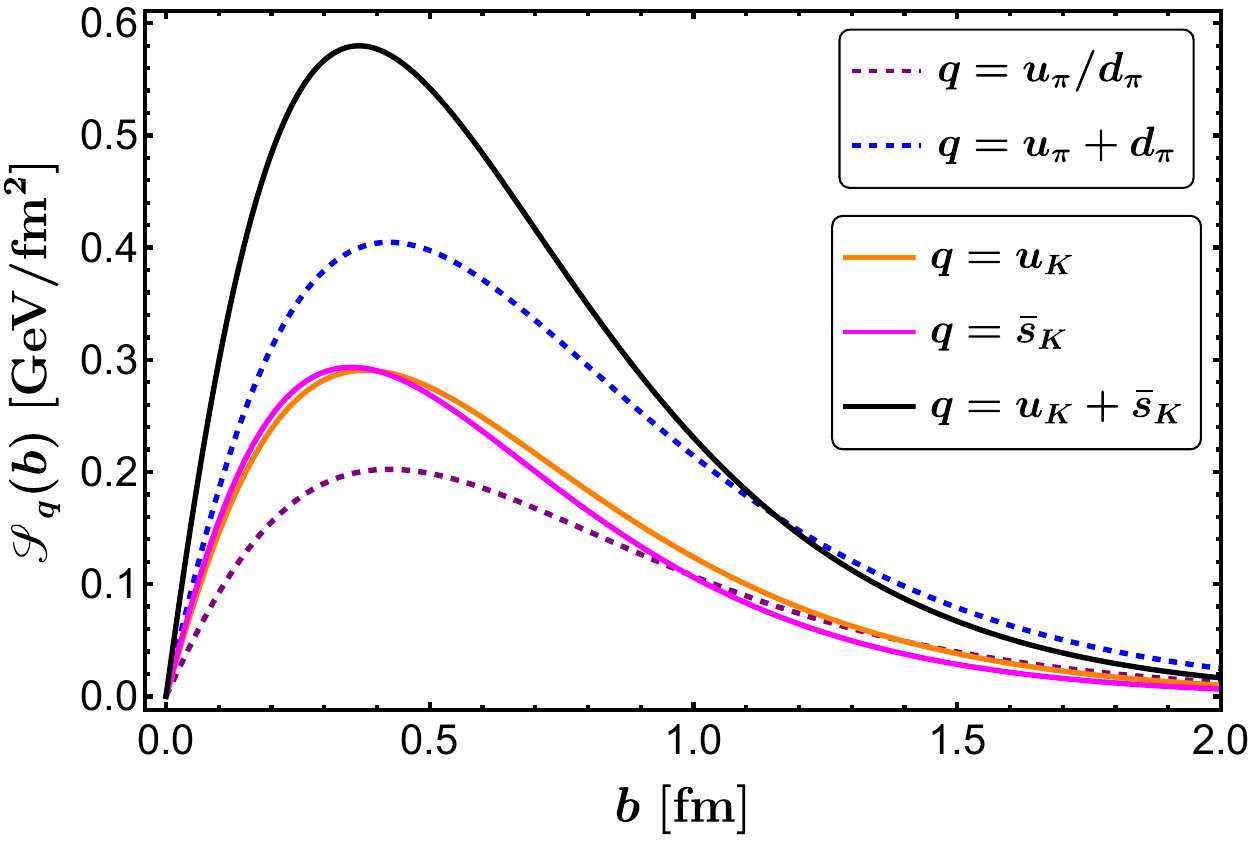}
\caption{Pion and kaon weighted transverse pressure and shear distributions and their quark flavor decompositions. The left panel shows the pressure distribution $\mathcal{P}_q(b)$, while the right panel shows the shear distribution $\mathcal{S}_q(b)$. The solid black curves denote the total kaon results, and the blue dashed curves represent the corresponding pion results. The orange and magenta curves show the $u$- and $\bar{s}$-quark contributions in the kaon, respectively. For the pion, the quark and antiquark contributions are identical and are shown by the purple dashed curves.}
	\label{fig:ps_constituents} 
\end{figure*}
\subsection{MECHANICAL DENSITIES AND RADII}\label{radii}
The GFF $D(Q^2)$ encodes information about the internal mechanical structure of hadrons. In particular, it determines the spatial distributions of pressure and shear force in the pion and kaon. These distributions are obtained from the two-dimensional Fourier transform of the $D(Q^2)$ with respect to the transverse momentum transfer, which maps the form factor from momentum space to impact-parameter space~\cite{Polyakov:2018zvc,Freese:2021czn}. They are given by
\begin{align}
	\label{Pressure and shear in terms of Dtilde}
	p(b)
	&=
	\frac{1}{2b}\frac{d}{db}\widetilde{D}(b)
	+\frac{1}{2}\frac{d^2}{db^2}\widetilde{D}(b)
	\nonumber\\
	&=
	\frac{1}{2b}\frac{d}{db}
	\left[
	b\frac{d}{db}\widetilde{D}(b)
	\right],
	\nonumber\\
	s(b)
	&=
	\frac{1}{b}\frac{d}{db}\widetilde{D}(b)
	-\frac{d^2}{db^2}\widetilde{D}(b)
	\nonumber\\
	&=
	-b\frac{d}{db}
	\left[
	\frac{1}{b}\frac{d}{db}\widetilde{D}(b)
	\right],
\end{align}
where $\widetilde{D}(b)$ denotes the Fourier transform of the $D(Q^2)$, with $b=|\vec b_\perp|$, defined as
\begin{align}
	\widetilde{D}(b)
	=
	\frac{1}{4P^+}
	\int \frac{d^2\vec q_\perp}{(2\pi)^2}
	e^{-i\vec q_\perp\cdot\vec b_\perp}
	D(-\vec q_\perp^{\,2}) \, .
\end{align}

The pressure distribution satisfies the von Laue stability condition~\cite{Polyakov:2018zvc,Freese:2021czn,Laue:1911lrk},
\begin{align}
	\int d^2 b\, p(b)=0 \, .
\end{align}
Equation~\eqref{Pressure and shear in terms of Dtilde} also shows that $p(b)$ and $s(b)$ are not independent. Their relation is encoded in the light-front equilibrium condition, analogous to the discussion near Eq.~(30) of Ref.~\cite{Polyakov:2018zvc},
\begin{align}
	p'(b)+\frac{1}{2}s'(b)+\frac{s(b)}{b}=0 \, .
\end{align}

Using Eq.~\eqref{Pressure and shear in terms of Dtilde} together with the recursion relations of the modified Bessel functions, one obtains the pressure and shear distributions for a dipole parametrization of the form factor $D(Q^2)$~\cite{Freese:2021czn},
\begin{align}
	\label{eqn:dipolepole:ps}
	p(b)
	&=
	\frac{D(0)m_D^4}{4P^+}
	\frac{1}{2^{3}\pi \Gamma(2)}
	\Big[
	(m_D b) K_{-1}(m_D b)-2K_0(m_D b)
	\Big] \, ,
	\nonumber\\[2mm]
	s(b)
	&=
	-\frac{D(0)m_D^4}{4P^+}
	\frac{1}{2^{2}\pi \Gamma(2)}
	(m_D b)K_{-1}(m_D b) \, ,
\end{align}
where the modified Bessel functions satisfy $K_{-1}(z)=K_1(z)$.

Figure~\ref{fig:mech_densities} shows the weighted pressure and shear distributions for the pion and kaon, $\mathcal{P}(b)=2\pi b\,P^+ p(b)$ and $\mathcal{S}(b)=2\pi b\,P^+ s(b)$. For both mesons, the pressure distribution exhibits a positive central region followed by a negative tail, consistent with the von Laue stability condition~\cite{Polyakov:2018zvc}. In the pion case, $\mathcal{P}(b)$ reaches its maximum at $b=0.11\,\mathrm{fm}$, changes sign at $b=0.49\,\mathrm{fm}$, and then gradually approaches zero at larger transverse distance. The kaon shows the same qualitative behavior, with a maximum at $b=0.09\,\mathrm{fm}$ and a zero crossing at the smaller distance $b=0.43\,\mathrm{fm}$, indicating a more compact pressure profile.

The weighted shear distribution remains positive for both mesons, as expected for a mechanically stable system~\cite{Polyakov:2018zvc}. It reaches its maximum at $b=0.42\,\mathrm{fm}$ for the pion and at $b=0.37\,\mathrm{fm}$ for the kaon, and then decreases toward zero at larger $b$. Taken together, the pressure and shear distributions indicate the expected pattern of short-range repulsion and intermediate-range attraction in the transverse plane.

Figure~\ref{fig:ps_constituents} displays the flavor decomposition of the weighted pressure and shear distributions. For both mesons, the flavor-separated pressure distributions retain the same qualitative structure as the total result, namely a positive core at small $b$ followed by a negative region at larger $b$. In the pion, the quark and antiquark contributions are identical. In the kaon, the $u$-quark pressure distribution reaches its maximum at $b=0.09\,\mathrm{fm}$ and crosses zero at $b=0.44\,\mathrm{fm}$, whereas the $\bar{s}$ contribution peaks at $b=0.09\,\mathrm{fm}$ and changes sign at $b=0.41\,\mathrm{fm}$. The corresponding weighted shear distributions peak at $b=0.38\,\mathrm{fm}$ for the $u$ quark and at $b=0.35\,\mathrm{fm}$ for the $\bar{s}$ quark.

The light-front mass (matter)\footnote{It is also referred to as the square of momentum ($P^+$) radius, as defined in Ref.~\cite{Freese:2021czn}.} radius squared is defined by
\begin{align}
	\label{eqn:radius:p+}
	\langle b_\perp^2 \rangle_{\mathrm{mat}}
	&\equiv
	\frac{1}{P^+}
	\int d^2 b_\perp \,
	b_\perp^2 \,
	\varepsilon(b_\perp)
	=
	4 \frac{dA(Q^2)}{dQ^2}\bigg|_{Q^2=0} \,,
\end{align}
where $\varepsilon(b_\perp)$ is the light-front momentum density,
\begin{align}
	\varepsilon(b_\perp)
	=
	P^+ \int \frac{d^{2}\vec{q}_{\perp}}{(2\pi)^{2}}
	e^{-i\vec{q}_{\perp}\cdot\vec{b}_\perp}
	A(-\vec{q}^{\,2}_{\perp}) \, .
\end{align}
Using the dipole form introduced above, this reduces to~\cite{Freese:2021czn}
\begin{align}
	\langle b_\perp^2 \rangle_{\mathrm{mat}}
	=
	\frac{8}{m_A^2} \, .
\end{align}

The two-dimensional mechanical radius of a spin-zero hadron is defined as~\cite{Freese:2021czn,Kim:2021jjf,Sain:2025kup}
\begin{align}
	\label{eqn:radius:mechanical}
	\langle b_\perp^2 \rangle_{\mathrm{mech}}
	&=
	\frac{
		\int d^2 b_\perp \,
		b_\perp^2
		\left[
		p(b_\perp) + \frac{1}{2} s(b_\perp)
		\right]
	}{
		\int d^2 b_\perp \,
		\left[
		p(b_\perp) + \frac{1}{2} s(b_\perp)
		\right]
	}
	\nonumber\\
	&=
	\frac{
		4D(0)
	}{
		\int_{-\infty}^0 dt \, D(t=-Q^2)
	} \,,
\end{align}
which is the light-front analogue of Eq.~(41) of Ref.~\cite{Polyakov:2018zvc}. With the same dipole parametrization, one obtains~\cite{Freese:2021czn}
\begin{align}
	\langle b_\perp^2 \rangle_{\mathrm{mech}}
	=
	\frac{4}{m_D^2} \, .
\end{align}

Using the fit parameters from the $L_{\max}=32$ results in Table~\ref{tab:dipole_fit}, we obtain for the pion
${\langle b_\perp^2 \rangle_{\mathrm{mat}}}=(0.35\pm0.04\,\mathrm{fm})^2$
and
${\langle b_\perp^2 \rangle_{\mathrm{mech}}}=(0.64\pm 0.06\,\mathrm{fm})^2$.
For the kaon, the corresponding values are
${\langle b_\perp^2 \rangle_{\mathrm{mat}}}=(0.31\pm 0.03\,\mathrm{fm})^2$
and
${\langle b_\perp^2 \rangle_{\mathrm{mech}}}=(0.55\pm 0.06\,\mathrm{fm})^2$.

The corresponding 3D mechanical radius of the pion in the Breit frame is  
\begin{equation}
r_{\pi,\rm mech}^2 \equiv \frac{3}{2} \langle b_\perp^2 \rangle_{\rm mech} = (0.78\pm 0.08~\rm{fm})^2,
\end{equation}
which is larger than the predicted pion matter (mass) radius,  
\begin{equation}
r_{\pi,\rm mat}^2 \equiv \frac{3}{2} \langle b_\perp^2 \rangle_{\rm mat} = (0.43\pm 0.04~\rm{fm})^2.
\end{equation}
The pion mass (matter) radius is in excellent agreement with recent lattice simulations~\cite{Hackett:2023nkr}, $r_{\pi,\rm mat}^2 = (0.41 \pm 0.01~\rm{fm})^2$, and reasonably consistent with the phenomenologically extracted value from the two-photon process $\gamma\gamma \to \pi^0\pi^0$~\cite{Kumano:2017lhr}, $r_{\pi,\rm mat}^2 = (0.32$--$0.39~\rm{fm})^2$, within uncertainties. Our prediction for the mechanical radius is larger than the lattice QCD result, $r_{\pi,\rm mech}^2 = (0.61 \pm 0.07~\rm{fm})^2$~\cite{Hackett:2023nkr}, but closer to the value obtained from the two-photon process, $r_{\rm mech}^2 = (0.82$--$0.88~\rm{fm})^2$~\cite{Kumano:2017lhr}.

We obtain the 3D mass and mechanical radii of the kaon in the Breit frame: 
\begin{equation}
\begin{aligned}
r_{K,{\rm mat}}^2 &= (0.38\pm 0.04~{\rm fm})^2, 
\\
r_{K,{\rm mech}}^2 &= (0.67\pm 0.07~\rm{fm})^2,
\end{aligned}
\end{equation}
respectively. As expected, the kaon's mass and mechanical radii are smaller than those of the pion.

\section{CONCLUSIONS}\label{sumary}
In this work, we have presented a study of the gravitational form factors $A(Q^2)$ and $D(Q^2)$ of the pion and kaon, along with their corresponding mechanical properties, within the BLFQ--NJL framework for light mesons. This approach combines light-front holography, longitudinal confinement, and color-singlet Nambu--Jona-Lasinio interactions~\cite{Jia:2018ary}. Solving the corresponding light-front Hamiltonian yields the valence light-front wave functions, which are subsequently used to evaluate the GFFs and the associated spatial distributions.

The form factor $A(Q^2)$, extracted from the $T^{++}$ component of the energy--momentum tensor (EMT), is in overall agreement with available lattice QCD and dispersive results for both the pion and the kaon. In contrast, the form factor $D(Q^2)$, extracted from the transverse components of the EMT, exhibits a substantially larger magnitude at low $Q^2$ compared to existing lattice QCD and dispersive studies. To characterize its low-momentum-transfer behavior, we employ a dipole parametrization. This enhancement of the $D$-term reflects a limitation of the current BLFQ--NJL framework, which is restricted to the leading valence sector and is expected to be more sensitive in the small-$x$ region relevant for the extraction of $D(Q^2)$. Nevertheless, at large $Q^2$ ($>0.5~{\rm GeV^2}$ for the pion and $>1~{\rm GeV^2}$ for the kaon), our results for $D(Q^2)$ are consistent with both lattice QCD and dispersive analyses.

Using the extracted form factors, we studied the pressure distribution $p(b)$ and the shear-force distribution $s(b)$ for both mesons. In each case, the pressure exhibits a positive core and a negative tail, consistent with the von Laue stability condition, while the shear distribution remains positive throughout the transverse plane. We also extracted the momentum and mechanical radii. For the pion, we find $r_{\pi,\rm mech}^2 = (0.78\pm 0.08~\rm{fm})^2$, which is close to the mechanical radius $r_{\rm mech}^2 = (0.82$--$0.88~\rm{fm})^2$ obtained from the two-photon process $\gamma\gamma \to \pi^0\pi^0$~\cite{Kumano:2017lhr}, while $r_{\pi,\rm mat}^2 = (0.43\pm 0.04~\rm{fm})^2$ is consistent with the lattice QCD prediction, $r_{\pi,\rm mat}^2 = (0.41 \pm 0.01~\rm{fm})^2$~\cite{Hackett:2023nkr}. For the kaon, the corresponding values are $r_{K,{\rm mech}}^2 = (0.67\pm 0.07~\rm{fm})^2$ and $r_{K,{\rm mat}}^2 = (0.38\pm 0.04~{\rm fm})^2$.

Overall, these results provide a quantitative picture of the gravitational structure and mechanical properties of light mesons within a nonperturbative light-front framework, while highlighting the need for a more complete treatment of contributions affecting the low-$Q^2$ behavior of the $D$-term.
\begin{acknowledgments}
We thank Zhimin Zhu, Jiangshan Lan, Jiatong Wu, and Tianyang Hu for many helpful discussions. 
This work is supported by the National Natural Science Foundation of China under Grant No.~12375143 and No.~12305095, by the Gansu International Collaboration and Talents Recruitment Base of Particle Physics (2023-2027), by the Senior Scientist Program funded by Gansu Province, Grant No.~25RCKA008.
X. Zhao is supported by Key Research Program of Frontier Sciences, Chinese Academy of Sciences, Grant No.~ZDBS-LY-7020, by the Foundation for Key Talents of Gansu Province, by the Central Funds Guiding the Local Science and Technology Development of Gansu Province, Grant No.~22ZY1QA006, by international partnership program of the Chinese Academy of Sciences, Grant No.~016GJHZ2022103FN, and by the Strategic Priority Research Program of the Chinese Academy of Sciences, Grant No.~XDB34000000.
J. P. Vary is supported by the Department of Energy under Grant No.~DE-SC0023692.  A portion of the computational resources were also provided by Advanced Computing Center in Taiyuan and by Sugon Computing Center in Xi’an.

\end{acknowledgments}

\bibliography{ref_updated.bib}

\end{document}